\documentstyle[prb,aps,multicol,epsf]{revtex}
\input psfig
\def\di{\displaystyle}
\def\psiuppl{\psi^{+}_{\uparrow}}
\def\psiup{\psi_{\uparrow}}
\def\psidnpl{\psi^{+}_{\downarrow}}
\def\psidn{\psi_{\downarrow}}
\def\de{\delta}
\def\De{\Delta}
\def\Db{\overline{\Delta}}
\def\zb{\overline{z}}
\def\Dp{\overline{\Delta}'}
\def\zp{\overline{z}'}
\def\om{\omega}
\def\omb{\overline{\omega}}
\def\Om{\Omega}
\def\ep{\varepsilon}
\def\la{\lambda}
\def\apb{\overline{\alpha}_{+}}
\def\amb{\overline{\alpha}_{-}}
\def\apm{\alpha_{\pm}}
\def\bpm{\beta_{\pm}}
\def\apmb{\overline{\alpha}_{\pm}}
\def\ap{\alpha_{+}}
\def\am{\alpha_{-}}
\def\bp{\beta_{+}}
\def\bm{\beta_{-}}
\def\Ga{\Gamma}
\def\ta{\tau_1\sigma_3}
\def\tb{\tau_2\sigma_3}
\def\tc{\tau_3\sigma_0}
\def\pa{\partial}
\def\Wb{\overline{W}}
\def\Wp{\overline{W}'}
\def\sqr#1#2{{\vcenter{\vbox{\hrule height .#2pt
      \hbox{\vrule width .#2pt height#1pt \kern#1pt
      \vrule width.#2pt}
      \hrule height.#2pt}}}}
\def\square{\mathchoice\sqr34\sqr34\sqr{2.1}3\sqr{1.5}3}

\begin{document}
\bibliographystyle{simpl1}

\title{Effect of Magnetic Impurities on Suppression of the
Transition Temperature in Disordered Superconductors}

\author{Robert A Smith}

\address{School of Physics and Astronomy, University of Birmingham,
Edgbaston, Birmingham B15 2TT, England\hfil\break
and Laboratory of Atomic and Solid State Physics,
Cornell University, Ithaca, New York 14853}

\author{Vinay Ambegaokar}

\address{Laboratory of Atomic and Solid State Physics,
Cornell University, Ithaca, New York 14853}

\maketitle
\bigskip

\begin{abstract}
We calculate the first-order perturbative correction to the transition
temperature $T_c$ in a superconductor with both non-magnetic and
magnetic impurities. We do this by first evaluating 
the correction to the effective
potential, $\Om(\De)$, and then obtain the first-order correction to
order parameter, $\De$, by finding the minimum of $\Om(\De)$.
Setting $\De=0$
finally enables $T_c$ to be evaluated. $T_c$ is now a function of both
the resistance per square, $R_{\square}$, a measure of the non-magnetic
disorder, and the spin-flip scattering rate, $1/\tau_s$, a measure of
magnetic disorder. We find that the effective pair-breaking rate per
magnetic impurity is virtually independent of the resistance per square
of the film, in agreement with an experiment of Chervenak and Valles.
This conclusion is supported both by the perturbative calculation, and by
a non-perturbative re-summation technique.
\end{abstract}

\section{Introduction}

Many experiments\cite{Fink94} performed on homogeneous disordered thin film
superconductors have shown that superconductivity is suppressed
by increasing disorder, as measured by the normal state resistance
per square, $R_{\square}$. The majority of the data is of the
transition temperature as a function of the resistance per square, 
$T_c(R_{\square})$\cite{RLCM,GB,HLG,LK}, although there is some data on the
upper critical field, $H_{c2}(T,R_{\square})$\cite{GB,HP,OKOK}, 
and the order parameter $\De(R_{\square})$\cite{VDG,Vall94}. The main data
to be explained thus consists of $T_c(R_{\square})$ curves for
different materials. To see how disorder might affect $T_c$, consider
the mean field equation,
\begin{equation}
T_{c0}=1.13\om_D\exp{\left[-{1\over N(0)(\la-\mu^*)}\right]},
\end{equation}
where $\la$ is the attractive BCS interaction mediated by
phonons of energy less than the Debye frequency, $\om_D$,
$\mu^*$ is the Coulomb pseudopotential, the effective strength of
the Coulomb repulsion, and $N(0)$ is the single particle density
of states at the Fermi surface. Obviously disorder could affect $T_c$
by changing $\la$, $\mu^*$, and $N(0)$. In particular the diffusive
motion of electrons caused by the disorder is known to lead to an
increased effective strength of the Coulomb interaction, as the screening
is less efficient than with ballistic electrons, and this leads to
an increase in $\mu^*$, and a decrease in $N(0)$. Calculating the 
first-order perturbative correction caused by the disorder shows that
we must consider all these processes together\cite{Fink87,SRW}. 
This is because the
disorder-screened Coulomb interaction has a low-momentum singularity
which leads to the separate effects being large; however, when they are
added together this singularity is cancelled, and the actual effect is 
much smaller than might be naively expected. The final result has the form
\begin{equation}
\ln{\left({T_c\over T_{c0}}\right)}=-{1\over 3}{R_{\square}\over R_0}
\ln^3{\left({1\over 2\pi T_{c0}\tau}\right)},
\end{equation}
where $R_0=2\pi h/e^2\approx 162k\Omega$ and $\tau$ is the elastic 
scattering time. We see that this curve is essentially ``universal'',
depending on only a single fitting parameter, 
$\beta=\ln{(1/2\pi T_{c0}\tau)}$. Experimentally it is found that
$T_c(R_{\square})$ curves from a wide variety of materials fit well
to this equation, or extensions of it that allow for stronger disorder.
The simplest extension simply consists of replacing $T_{c0}$ by $T_c$
on the right-hand side of Eq. (2), which leads to the cubic equation
\begin{equation}
x={t\over 3}(\beta+x)^3,
\end{equation}
where $x=\ln{(T_{c0}/T_c)}$ and $t=R_{\square}/R_0$. This equation shows
unphysical reentrance at strong disorder, an artefact which is removed
by either a renormalization group treatment\cite{Fink87}, or the use of 
a non-perturbative resummation technique\cite{OF} to yield the formula
\begin{equation}
\ln{\left({T_c\over T_{c0}}\right)}={1\over\lambda}
-{1\over 2\sqrt{t}}\ln{\left({1+\sqrt{t}/\lambda\over
1-\sqrt{t}/\lambda}\right)}.
\end{equation}
\par
The fact that most data can be fit to a single curve is pleasing in
that it shows that the basic ingredients of our theory -- disorder,
BCS attraction and Coulomb repulsion -- are correct. However it does
not allow analysis of the sensitivity of the theory or experiment to  
changes in details of the system, such as the exact form of the 
phonon-mediated attraction. Moreover there are other theories\cite{Belitz}
which posit the importance of such details which give predictions that are
equally in agreement with experiment. We see that the $T_c(R_{\square})$
curves alone are not enough to allow consideration of the relative
merits of different theories.
What we would like to do is to add some additional parameter
to the experimental system to give a whole new set of data -- for
example a family of $T_c(R_{\square})$ curves for a single material
as this new parameter is altered. Chervenak and Valles\cite{CV} have recently
performed an experiment of this type in which magnetic $Gd$ impurities
are added to thin films of $Pb_{0.9}Bi_{0.1}$. This introduces the
new feature of spin-flip scattering to the system, which is measured
by the spin-flip scattering rate, $1/\tau_s$. The task of the theorist
is to now make predictions for $T_c$ as a function of
both $R_{\square}$ (the measure of non-magnetic disorder), and $1/\tau_s$,
(the measure of magnetic disorder), and to compare these to experiment.
\par
In this paper we calculate the first-order perturbative correction to
the transition temperature, $T_c$, of a superconductor with both
non-magnetic and magnetic impurities. The model used consists of a
featureless BCS attraction, $-\lambda$, and a Coulomb repulsion, $V_C(q)$,
between electrons which scatter off non-magnetic and magnetic impurities.
The model is the simplest one that contains the essential physics, and
its shortcoming of not considering the details of the attractive
interaction is offset by the fact that we can consider all processes to
a given order of perturbation theory. This is an important consideration
in view of the cancellation of low-momentum singularities in the screened
Coulomb potential discussed in the opening paragraph. In fact, an obvious
question is whether this cancellation persists in the presence of magnetic 
impurities. We find that this is indeed the case, and so the details of
the screened Coulomb interaction are removed, leading to a ``universal''
form for $T_c(R_{\square},1/\tau_s)$. 
\par
The main result of the paper is that
the pair-breaking rate per magnetic impurity, $\alpha'(R_{\square})$,
defined by 
\begin{equation}
\alpha'(R_{\square})={T_c(R_{\square},0)-T_c(R_{\square},1/\tau_s)
\over 1/\tau_s}
\end{equation}
is roughly independent of $R_{\square}$ except near the
superconductor-insulator transition, in agreement with experiment.
This is confirmed both by first-order perturbation theory, and also
by a non-perturbative resummation technique which we introduce to 
remove concerns about reentrance problems at stronger disorder. This
agreement of the two theoretical approaches with each other and the
experimental data gives us confidence in our results.
\par
To calculate the correction to $T_c$ we use a collective mode formalism
derived in a previous paper\cite{SRW} (which we refer to as {\bf I} from now
on) on the suppression of $\De$ by non-magnetic disorder.
The introduction of
magnetic impurities means that we have to modify the formalism
somewhat, and so we include most of the derivation in this paper.
The method used in this paper to evaluate the correction to $T_c$
proceeds in three stages. First we find the first-order correction
to the grand canonical potential, $\Om_1(\De)$, of the superconductor
due to fluctuations of its collective modes. Then by minimizing the
total grand canonical potential, $\Om_0(\De)+\Om_1(\De)$ with respect
to the order parameter, $\De$, we obtain the first-order correction to the
order parameter self-consistency equation. Finally by setting $\De=0$
we obtain the first-order correction to transition temperature $T_c$. 
The method we use has
the advantage that it is impossible to ``miss diagrams'' since there
is only one diagram in the $\Om_1(\De)$ calculation, and we also obtain
the equation for $\De$ at no extra cost. The equations for $T_c$ and
$\De$ must reduce to those of {\bf I} when we set spin-flip scattering
to zero, providing a useful consistency check.
A key result of the calculation in {\bf I} was that the
singularity in the screened Coulomb potential persists below $T_c$,
and that this singularity is cancelled in the formula for the
suppression of $\De$ in a similar manner to the cancellation in the
formula for $T_c$. Therefore an important question
is whether this singularity and cancellation remains when magnetic
impurities are added. We show that this is indeed the case, and moreover
that the cancellation is due to gauge invariance, and will occur in
first-order perturbation theory in the presence of any kind of impurity
scattering. In other words, it is not possible to obtain stronger suppression
of the transition temperature by introducing some exotic scattering
mechanism. It is also reassuring to know that an otherwise mysterious
cancellation between diagrams has its physical origin in gauge invariance,
and we hope that similar arguments may be applied to show that the
result holds to all orders in perturbation theory.
\par
The outline of the rest of the paper is as follows. In section
\uppercase\expandafter{\romannumeral2} we derive the matrix formalism
for superconductors with magnetic impurities, and the collective
mode approach we will use. We derive the RPA screened bosonic
propagators, and show that low-momentum singularities persist in
the screened Coulomb propagator below $T_c$. In section
\uppercase\expandafter{\romannumeral3} we derive the first order
perturbative correction to the grand canonical potential
$\Om_1(\De)$, and from this the correction to the order parameter $\De$.
In section \uppercase\expandafter{\romannumeral4} we set $\De=0$ to
obtain the correction to transition temperature $T_c$. In section
\uppercase\expandafter{\romannumeral5} we calculate $T_c$ numerically
using both the perturbative results of section 
\uppercase\expandafter{\romannumeral4} and a recently developed
non-pertubative technique, and compare to experiment.

\section{Derivation of the $4\times 4$ Matrix Formalism}

\noindent{\bf Superconductivity with Magnetic Impurities:}
\medskip

We consider a system of electrons that scatter off static non-magnetic
and magnetic impurities, and interact with each other via the long-range
Coulomb interaction and the BCS attraction. The scattering from static
impurities is described by Hamiltonian
\begin{equation}
H_{e-i} = \sum_{\alpha\beta}\int d{\bf x}\psi^{+}_{\alpha}({\bf x})
\left\{\left[-{\nabla^2_x\over 2m}+\sum_i u_0({\bf x}-{\bf x_i})\right]
\delta_{\alpha\beta} + \sum_j J({\bf x-x_j}){\bf S_j\cdot\sigma_{\alpha\beta}}
\right\}\psi_{\beta}({\bf x}),
\end{equation}
where $\psi^{+}_{\alpha}$, $\psi_{\alpha}$ are the electron creation
and annhilation operators, $u_0({\bf x}-{\bf x_i})$ is the
impurity potential at ${\bf x}$ due to a non-magnetic impurity at
${\bf x_i}$, ${\bf S_j}$ is a magnetic impurity spin moment at ${\bf x_j}$,
and $J({\bf x})$ is the electron-impurity exchange coupling. The potential
and spin-flip scattering rates are then given by
\begin{eqnarray}
{1\over\tau_0}&=&2\pi N(0)n_i |u_0|^2\nonumber\\
{1\over\tau_s}&=&2\pi N(0)n_j J^2 S(S+1),
\end{eqnarray}
where $u_0$ is the Fourier transform of $u_0({\bf x})$, and $J$ is the
Fourier transform of $J({\bf x})$, both assumed
independent of momentum, $n_i$ is the non-magnetic impurity density,
and $n_j$ the magnetic impurity density.
\par
The Coulomb repulsion between electrons is described by Hamiltonian
\begin{equation}
H_C=\sum_{\alpha\beta}\int d{\bf x} \int d{\bf x'}
\psi^{+}_{\alpha}({\bf x})\psi_{\alpha}({\bf x})
{e^2\over |{\bf x}-{\bf x'}|}
\psi^{+}_{\beta}({\bf x'})\psi_{\beta}({\bf x'}),
\end{equation}
leading to a bare Coulomb propagator that is just the Fourier transform
of the above potential.
\par
The BCS attraction is described by the Hamiltonian
\begin{equation}
H_{BCS}=-\lambda\sum_{\alpha\beta}\int d{\bf x}
\psi^{+}_{\alpha}({\bf x})\psi_{\alpha}({\bf x})
\psi^{+}_{\beta}({\bf x})\psi_{\beta}({\bf x}),
\end{equation}
corresponding to an instantaneous contact interaction
$-\lambda\delta({\bf x}-{\bf x'})$.

Having introduced the model Hamiltonian we need to describe the system,
we discuss the standard four-dimensional matrix representation\cite{Grif}
needed
to describe a superconductor with magnetic impurities. We need four
components to describe the two spin degrees of freedom, and the two
types of correlation -- the usual particle-hole correlation
$<\psiup\psiuppl>$, amd the anomalous pairing correlation
$<\psiup\psidn>$. We introduce
the four-dimensional vector operator
\begin{equation}
\Psi = \left(\matrix{\psiup\cr \psidn\cr \psidnpl\cr \psiuppl\cr}
         \right) \quad;\quad
  \Psi^{+}= \left(\matrix{\psiuppl&\psidnpl&\psidn&\psiup\cr}\right),
\end{equation}
with matrix propagator
\begin{equation}
<\Psi\Psi^{+}> = \left[\matrix{
<\psiup\psiuppl>&<\psiup\psidnpl>&<\psiup\psidn>&<\psiup\psiup>\cr
<\psidn\psiuppl>&<\psidn\psidnpl>&<\psidn\psidn>&<\psidn\psiup>\cr
<\psidnpl\psiuppl>&<\psidnpl\psidnpl>&<\psidnpl\psidn>&<\psidnpl\psiup>\cr
<\psiuppl\psiuppl>&<\psiuppl\psidnpl>&<\psiuppl\psidn>&<\psiuppl\psiup>\cr
}\right].
\end{equation}
In the normal state the temperature Green function is
\begin{equation}
G(k,i\om)={1\over z-\ep_k\tau_3\sigma_0},
\end{equation}
where $z=i\om$, $\om=(2n+1)\pi T$ is a Fermi Matsubara frequency,
and the $\tau_i$ and $\sigma_i$ are Pauli matrices operating on different
spaces. The $\sigma_i$ operate in the usual spin space, whilst the
$\tau_i$ operate in the Nambu (electron-hole) space. The diagrammatic
rules are then the same as in the normal state except for the matrix
structure of the Green function, and the presence of matrix
$\tau_3\sigma_0$ at each interaction or impurity vertex due to the
electron density operator being written in the form
\begin{equation}
\rho=\psiuppl\psiup+\psidnpl\psidn=
{1\over 2}\Psi^{+}\tau_3\sigma_0\Psi.
\end{equation}
\par
The pairing correlations in the clean superconductor can then be taken
into account self-consistently as shown in Fig. (1a). Making the ansatz
$\Sigma=\De\tau_1\sigma_3$ for the self-energy, the Green function for
the pure superconductor becomes
\begin{equation}
G_0(k,z)={1\over z-\ep_k\tau_3\sigma_0-\De\tau_1\sigma_3}
={z+\ep_k\tau_3\sigma_0+\De\tau_1\sigma_3\over z^2-\ep_k^2-\De^2},
\end{equation}
and the diagram of Fig. (1a) gives self-energy
\begin{eqnarray}
\Sigma\displaystyle&=&-\lambda T\sum_\om N(0)\int d\ep_k
{\tau_3\sigma_0(z+\ep_k\tau_3\sigma_0+\De\tau_1\sigma_3)\tau_3\sigma_0
\over \ep_k^2-z^2+\De^2}\nonumber\\
&=&\displaystyle N(0)\lambda\De\tau_1\sigma_3 
T\sum_\om {1\over\sqrt{\om^2+\De^2}},
\end{eqnarray}
which gives us the usual BCS self-consistency equation
\begin{equation}
1=N(0)\lambda T\sum_\omega {1\over\sqrt{\omega^2+\De^2}}.
\end{equation}

We can treat the presence of non-magnetic and magnetic impurities by
including an extra self-energy diagram to describe the dressing of the
electron line by impurities as shown in Fig. (1b). We then make the
ansatz that the pairing energy has the form $\Sigma_p=\De\tau_1\sigma_3$,
and the impurity self-energy has the form
$\Sigma_{imp}=-(\zb-z)+(\Db-\De)\tau_1\sigma_3$, so that the Green
function for the dirty superconductor is
\begin{equation}
G_0(k,z)={\zb+\ep_k\tau_3\sigma_0+\Db\tau_1\sigma_3\over
\zb^2-\ep_k^2-\Db^2},
\end{equation}
which is just the Green function for the clean superconductor with
$z$, $\De$, replaced by $\zb$, $\Db$, respectively. Since the
impurity line has the form
\begin{equation}
\label{imp}
\Ga_0={1\over 2\pi N(0)\tau_0}\tau_3\sigma_0\otimes\tau_3\sigma_0
+{1\over 6\pi N(0)\tau_s}\left[\tau_0\sigma_1\otimes\tau_0\sigma_1
+\tau_0\sigma_2\otimes\tau_0\sigma_2
+\tau_3\sigma_3\otimes\tau_3\sigma_3\right],
\end{equation}
we obtain the self-consistency equation for $\zb=i\omb$ and $\Db$,
\begin{equation}
\label{selfcons}
\omb-\om=\left({1\over 2\tau_0}+{1\over 2\tau_s}\right)
{\omb\over\sqrt{\omb^2+\Db^2}}\quad;\quad
\Db-\De=\left({1\over 2\tau_0}-{1\over 2\tau_s}\right)
{\Db\over\sqrt{\omb^2+\Db^2}}.
\end{equation}
The diagrammatic definition of the pairing energy, $\Sigma_p$, leads to
the same self-consistency equation for $\De$ as in the pure case except
that $\om$, $\De$, are replaced by $\omb$, $\Db$. In the absence of
magnetic impurities -- i.e. $1/\tau_s=0$ -- we see that
$\omb/\Db=\om/\De$, and the equation for $\De$ is unchanged. This is
Anderson's theorem\cite{PWA} 
that superconductivity is unaffected by non-magnetic
impurities at mean-field level. In the presence of magnetic impurities
we see that $\omb/\Db\ne\om/\De$, and if we define $u=\omb/\Db$,
$\zeta=1/\tau_s\De$, the problem reduces to solving the equation
\begin{equation}
\label{transc}
{\om\over\De} = u\left(1-\zeta{1\over\sqrt{u^2+1}}\right).
\end{equation}
The self-consistency equation for $\De$ then takes the form
\begin{equation}
\label{mfeqn}
1=N(0)\lambda T\sum_\om {1\over\sqrt{u^2+1}},
\end{equation}
and in particular if we set $\De=0$ we get for $T_c$,
\begin{equation}
1=N(0)\lambda T_c\sum_\om {1\over |\om|+1/\tau_s}.
\end{equation}
Subtracting off the equation for $T_{c0}$, the transition temperature
in the absence of magnetic impurities leads to the famous 
result\cite{AG}
\begin{equation}
\log{\left({T_c\over T_{c0}}\right)} =
\psi\left({1\over 2}\right) -
\psi\left({1\over 2}+{1\over 2\pi T_c\tau_s}\right).
\end{equation}
\null
\medskip
\noindent{\bf Collective Mode Formalism and RPA:}

The idea of the collective mode formalism is to treat the screened
interactions in the system as bosonic collective modes. The relevant
bosonic operators are order parameter amplitude and phase, and electron
density. These are the only modes that are coupled by a bare interaction
-- the BCS interaction for order parameter amplitude and phase,
the Coulomb interaction for electron density. The main advantage of this
approach is that we are able to treat order parameter fluctuations and
the Coulomb interaction on an identical footing.
This procedure may be formally carried out within the path integral
theory of superconductivity by decoupling the two four-fermion interaction
terms with the introduction of appropriate collective variables.
This is discussed
in detail in the paper of Eckern and Pelzer\cite{EP88}. The end result is that
there are three effective bosonic modes, order parameter amplitude and
phase and electronic density, and each can be written in the form
\begin{equation}
\hat{O}_i={1\over 2}\Psi^+ M_i\Psi,
\end{equation}
with matrices $M_{\Delta}=\tau_1\sigma_3$ for order parameter amplitude,
$M_{\phi}=\tau_2\sigma_3$ for order parameter phase, and
$M_{\rho}=\tau_3\sigma_0$ for electronic density. Interactions occur by
exchange of these collective modes, and so the effective interaction
potential is now a $3\times 3$ matrix. The screened interaction is found
from the equation
\begin{equation}
V_{ij}=V^0_{ij}+\sum_{kl}V^0_{ik}\Pi_{kl}V_{lj},
\end{equation}
as shown in Fig. (2). Here $\Pi_{ij}$ is the polarization operator
and $V^0_{ij}$ is the bare interaction matrix which is given by
\begin{equation}
V^0=\left(\matrix{-\lambda/2&0&0\cr 0&-\lambda/2&0\cr 
0&0&{4\pi e^2/q^2}\cr}\right),
\end{equation}
the BCS attraction being split equally between the two order parameter
modes. The only new diagrammatic feature is that any of the matrices
$M_i$ can now appear at an interaction vertex, corresponding to
interaction with the collective variable described by that matrix.

In order to carry out the calculations in section
\uppercase\expandafter{\romannumeral3}, we need the impurity dressed
RPA polarization bubbles, $\Pi_{ij}$, shown in Fig. (3). To evaluate the
polarization bubble $\Pi_{ij}$ we must first evaluate the geometric
series
\begin{equation}
\Pi=S+S\Ga_0 S+S\Ga_0 S\Ga_0 S+\dots,
\end{equation}
where
$\Ga_0$ is the impurity line defined in Eqn. ({\ref{imp}), 
and $S$ is the momentum sum of a direct product of Green functions
\begin{equation}
S=\sum_k G(k,i\om)\otimes G(k+q,i\om+i\Om).
\end{equation}
Since we do not need the complete matrix structure of $\Pi$, but just
its traces with two matrices from the set $\ta$, $\tb$, $\tc$, we
actually evaluate the impurity dressed vertices $\Pi_j$ which have one 
matrix from the above set inserted between the two terms of the direct 
product in $\Pi$. These satisfy the equations
\begin{equation}
\Pi_j=S_j+S\Ga_0\Pi_j,
\end{equation}
where $S_j$ is obtained by inserting the matrix $M_j$ between the two
terms of the direct product in $S$.
These equations are solved in Appendix A to obtain the results
\begin{eqnarray}
\Pi_{\De}&=&-{\pi N(0)\over D_+}
\left[{UU'+uu'-1\over UU'}-{i(u'+u)\over UU'}\ta\right]\ta\nonumber\\
\Pi_{\phi}&=&-{\pi N(0)\over D_-}
\left[{UU'+uu'+1\over UU'}-{i(u'-u)\over UU'}\ta\right]\tb\nonumber\\
\Pi_{\phi}&=&{\pi N(0)\over D_-}
\left[{UU'-uu'-1\over UU'}+{i(u'-u)\over UU'}\ta\right]\tc,
\end{eqnarray}
where $U=\sqrt{u^2+1}$, $u'=u(\omega+\Omega)=u(\omega')$,
$U'=\sqrt{u'^2+1}$ and
\begin{equation}
D_{\pm}=
\left[Dq^2+\De U+\De U'+{1\over\tau_s}
\left({uu'\mp 1\over UU'}-1\right)\right].
\end{equation}
We can finally obtain the non-zero polarization bubbles $\Pi_{ij}$
by inserting the second matrix from the set $\ta$, $\tb$, $\tc$ into
$\Pi_j$, taking the trace, and recalling the factor $-1$ for a fermion
loop. This yields
\begin{eqnarray}
\label{piall}
\Pi_{\De\De}(q,\Om) &=&\displaystyle\pi N(0)T\sum_\om \left[
{UU'+uu'-1\over UU'}\right]{1\over
\left(Dq^2+\De U+\De U'-\di{1\over\tau_s}\left[{UU'-uu'+1\over UU'}\right]
\right)}\nonumber\\
\Pi_{\phi\phi}(q,\Om) &=&\displaystyle \pi N(0)T\sum_\om \left[
{UU'+uu'+1\over UU'}\right]{1\over
\left(Dq^2+\De U+\De U'-\di{1\over\tau_s}\left[{UU'-uu'-1\over UU'}\right]
\right)}\nonumber\\
\Pi_{\rho\rho}(q,\Om) &=&\displaystyle-\pi N(0)T\sum_\om \left[
{UU'-uu'-1\over UU'}\right]{1\over
\left(Dq^2+\De U+\De U'-\di{1\over\tau_s}\left[{UU'-uu'-1\over UU'}\right]
\right)} + N(0)\nonumber\\
\Pi_{\phi\rho}(q,\Om) &=&\displaystyle-\pi N(0)T\sum_\om \left[
{u'-u\over UU'}\right]{1\over
\left(Dq^2+\De U+\De U'-\di{1\over\tau_s}\left[{UU'-uu'-1\over UU'}\right]
\right)}=-\Pi_{\rho\phi}(q,\Om).
\end{eqnarray}
We note that if we set $1/\tau_s=0$ in Eqn. (\ref{piall}) we will obtain
exactly the results found in {\bf I}, as of course we must. The screened
potentials $V_{ij}$ are then given by
\begin{equation}
\label{Veqn}
V=\left[\matrix{
(-\la^{-1}+\Pi_{\De\De})^{-1}&0&0\cr
0&[(2V_C(q))^{-1}+\Pi_{\rho\rho}]/{\cal{D}}&-\Pi_{\phi\rho}/{\cal{D}}\cr
0&\Pi_{\phi\rho}/{\cal{D}}&(-\la^{-1}+\Pi_{\phi\phi})/{\cal{D}}\cr}\right],
\end{equation}
where
\begin{equation}
{\cal{D}}\equiv(-\la^{-1}+\Pi_{\phi\phi})[(2V_C(q))^{-1}+\Pi_{\rho\rho}]
+\Pi_{\phi\rho}^2.
\end{equation}
The coupling between phase and density fluctuations caused by the non-zero
value of $\Pi_{\phi\rho}=-\Pi_{\rho\phi}$ is a manifestation
of gauge invariance.

We can next show that the propagators $V_{\phi\phi}$, $V_{\phi\rho}$
and $V_{\rho\rho}$ all have a low-momentum singularity of the form
$1/q^{d-1}$ for all non-zero frequencies and all temperatures. In other
words, these propagators have the same long-range behavior as the unscreened
Coulomb potential. An analogous situation occurs for the 
disorder screened potential in the normal metal, where the singularity 
is known to
strongly affect the properties of the system. To show the existence of
the singularity we simply need to show that the denominator ${\cal{D}}$
vanishes at $q=0$ for all $\Om\ne 0$ and all $T$. Since
$V_C(q)\sim 1/q^{d-1}$, we need only prove that
\begin{equation}
[-\lambda^{-1}\Pi_{\phi\phi}(0,\Om)]\Pi_{\rho\rho}(0,\Om)+
\Pi_{\phi\rho}(0,\Om)^2 = 0.
\end{equation}
This is proved in Appendix B where we show that
\begin{equation}
[-\lambda^{-1}+\Pi_{\phi\phi}(0,\Om)]={\Om\over 2\De}\Pi_{\phi\rho}(0,\Om)
\quad,\quad \Pi_{\phi\rho}(0,\Om)=-{\Om\over\De}\Pi_{\rho\rho}(0,\Om),
\end{equation}
by direct calculation. We also show that this result is guaranteed
by gauge invariance and is therefore true no matter which scattering
mechanisms we include.

\section{First Order Correction to Grand Potential and Order-Parameter
Self-Consistency Equation}

In this section we evaluate the first-order perturbation correction to
the grand potential, $\Om_1(\De)$. By minimising the sum
$\Om_0(\De)+\Om_1(\De)$ with respect to $\De$ we obtain the corresponding
correction to the order parameter self-consistency equation.
This method was first used for the system with only non-magnetic
impurities by Eckern and Pelzer\cite{EP88}, and we choose to use it as it
involves the smallest number of diagrams. The same result for $T_c$
can also be obtained using the Eliashberg diagrams for $\De$ shown
in Fig. (5), or the pair propagator diagrams shown in Fig. (6).
\par
The diagram for the first order correction to the grand potential
simply consists of the ``string of bubbles'' diagram shown in Fig. (4)
Since the polarization bubbles in this diagram are just those
evaluated in the previous section, we have all the information we
need to derive $\Om_1(\De)$. The only thing we need to remember is
the extra symmetry factor of $1/n$ required for the diagram with
$n$ bubbles. So whereas previously the RPA equation involved the series
\begin{equation}
V = V_0+V_0\Pi V_0+V_0\Pi V_0\Pi V_0+\dots
= (V_0^{-1}-\Pi)^{-1},
\end{equation}
it now becomes
\begin{equation}
\Om_1 = V_0\Pi+{1\over 2}V_0\Pi V_0\Pi+{1\over 3}V_0\Pi V_0\Pi V_0\Pi
+\dots= -\log{[V_0^{-1}-\Pi]}+\log{[V_0^{-1}]}.
\end{equation}
After summing over all the internal variables -- momentum $q$, Bose
Matsubara frequency $\Om$ and the three bosonic modes -- we end up
with the final expression for $\Om_1$,
\begin{equation}
\label{gpcorr}
\Om_1=-T\sum_\Om\sum_q\left\{\log{(-\la^{-1}+\Pi_{\De\De}(q,\Om))}
+\log{\big[(-\la^{-1}+\Pi_{\phi\phi}(q,\Om))
[(2V_C(q))^{-1}+\Pi_{\rho\rho}(q,\Om)]+
\Pi_{\phi\rho}(q,\Om)^2\big]}\right\}.
\end{equation}
To proceed we need to minimise the total grand potential,
\begin{equation}
\label{minpot}
{\pa\over\pa\De}\left\{\Om_0(\De)+\Om_1(\De)\right\}=0.
\end{equation}
The mean-field grand potential, $\Om_0(\De)$, is given by
\begin{equation}
\Om_0(\De)=N(0){\De^2\over\la}+T\sum_\om\sum_k
\hbox{Tr}\left[\log{(\zb-\xi_k\tc-\Db\ta)}\right],
\end{equation}
and after taking the derivative of $\Om_0(\De)$ with respect to $\De$,
we see that Eqn. (\ref{minpot}) takes the form
\begin{equation}
{1\over\la}-\pi N(0)T\sum_\om {1\over U} = -{1\over\De}
{\pa\Om_1\over\pa\De}.
\end{equation}
The next step in the procedure is to evaluate $\pa\Om_1(\De)/\pa\De$.
From Eqn. (\ref{gpcorr}) we see that $\pa/\pa\De$ will be acting upon the
$\Pi_{ij}$ to give
\begin{equation}
\label{newsc}
-{\pa\Om_1\over\pa\De} = T\sum_\Om \sum_q \left\{
{\pa\Pi_{\De\De}\over\pa\De}V_{\De\De}+
{\pa\Pi_{\phi\phi}\over\pa\De}V_{\phi\phi}-
2{\pa\Pi_{\phi\rho}\over\pa\De}V_{\phi\rho}+
{\pa\Pi_{\rho\rho}\over\pa\De}V_{\rho\rho}\right\}.
\end{equation}
From Eqn. (\ref{piall}) we see that the $\pa/\pa\De$ can act either on the
coherence factor or on the denominator in the expression for $\Pi_{ij}$.
Acting on the denominator gives a result proportional to the
denominator squared, corresponding to the two-ladder diagrams of
Fig. (5a) in the Eliashberg approach. Similarly acting on the coherence
factor leads to the one-ladder diagrams of Fig. (5b). We note that the
explicit evaluation of the three-ladder diagrams of Fig. (5c) will give
a zero result.
\par
The only difficulty in taking the derivatives of the polarization
bubbles, $\Pi_{ij}$, with respect to $\De$ is that the quantity
$u(\om)$ present in all these equations satisfies the transcendental
Eqn. (\ref{transc}). In Appendix C we evaluate the derivatives of
the $\Pi_{ij}$ to obtain the results
\begin{eqnarray}
{\pa\Pi_{\De\De}\over\pa\De}&=&-2\pi N(0)T\sum_\om \left\{
\left[1-{\zeta\over U^3}\right]^{-1}\times 
{1\over\De}{u(u'+u)\over U^3U'}{1\over D_{+}}+
\left(1+{uu'-1\over UU'}\right)\left\{{1\over U}-{\zeta\over U^2}
\left(1+{u(u'-u)\over UU'}\right)\right\}{1\over D_{+}^2}\right\}
\nonumber\\
{\pa\Pi_{\phi\phi}\over\pa\De} &=&-2\pi N(0)T\sum_\om \left\{
\left[1-{\zeta\over U^3}\right]^{-1}\times 
{1\over\De}{u(u'-u)\over U^3U'}{1\over D_{-}}+
\left(1+{uu'+1\over UU'}\right)\left\{{1\over U}-{\zeta\over U^2}
\left(1+{u(u'+u)\over UU'}\right)\right\}{1\over D_{-}^2}\right\}
\nonumber\\
{\pa\Pi_{\rho\rho}\over\pa\De} &=&-2\pi N(0)T\sum_\om \left\{
\left[1-{\zeta\over U^3}\right]^{-1}\times
{1\over\De}{u(u'-u)\over U^3U'}{1\over D_{-}}-
\left(1-{uu'+1\over UU'}\right)\left\{{1\over U}-{\zeta\over U^2}
\left(1+{u(u'+u)\over UU'}\right)\right\}{1\over D_{-}^2}\right\}
\nonumber\\
{\pa\Pi_{\phi\rho}\over\pa\De} &=& -2\pi N(0)T\sum_\om \left\{
\left[1-{\zeta\over U^3}\right]^{-1}\times 
{1\over\De}{u(uu'+1)\over U^3U'}{1\over D_{-}}-
\left({u'-u\over UU'}\right)\left\{{1\over U}-{\zeta\over U^2}
\left(1+{u(u'+u)\over UU'}\right)\right\}{1\over D_{-}^2}\right\}.
\end{eqnarray}
These formulas together with Eqn. (\ref{newsc}) lead to our final result for
the first order correction to the order parameter self-consistency
equation:
\begin{eqnarray}
\label{finalsc}
&&\di{1\over N(0)\la}-\pi T\sum_\om {1\over U}={\pi\over 2}T\sum_\om
T\sum_\Om \sum_q \left[1-{\zeta\over U^3}\right]^{-1}\times\nonumber\\
&&\di\left\{\left[{1\over\De^2}{1\over D_{+}}{u(u'+u)\over U^3U'}
+{1\over\De}{1\over D_{+}^2}{1\over U}\left\{1-{\zeta\over U}\left(
1+{u(u'-u)\over UU'}\right)\right\}\left(1+{uu'-1\over UU'}\right)
\right]V_{\De\De}(q,\Om)\right.\nonumber\\
&&\di+{1\over\De^2}{1\over D_{-}}
\left[{u(u'-u)\over U^3U'}V_{\phi\phi}(q,\Om)-
{2u(uu'+1)\over U^3U'}V_{\phi\rho}(q,\Om)+
{u(u'-u)\over U^3U'}V_{\rho\rho}(q,\Om)\right]\nonumber\\
&&\di\left.+{1\over\De}{1\over D_{-}^2}\left\{1-{\zeta\over U}
\left(1+{u(u'-u)\over UU'}\right)\right\}
\left[\left(1+{uu'+1\over UU'}\right)V_{\phi\phi}(q,\Om)+
{2(u'-u)\over UU'}V_{\phi\rho}(q,\Om)-
\left(1-{uu'+1\over UU'}\right)V_{\rho\rho}\right]\right\}.
\end{eqnarray}
\par
The above formula is valid for all temperatures $0\le T\le T_c$, 
but we are usually interested
in the special cases $T=0$ and $\De=0$ (i.e. $T=T_c$). In these two
cases the sum over $\om$ on the LHS can be performed analytically to yield
the two simple forms
\begin{eqnarray}
\log{\left({\De(0)\over\De_0(0)}\right)} &=&{\pi\over 2}
T\sum_\Om \sum_q T\sum_\om \dots \nonumber\\
\log{\left({T_c\over T_{c0}}\right)} &=&{\pi\over 2}
T\sum_\Om \sum_q T\sum_\om \dots
\end{eqnarray}
\par
Having noted the presence of the $1/q^{d-1}$ singularities in the
potentials $V_{\phi\phi}$, $V_{\phi\rho}$ and $V_{\rho\rho}$, we should
now see whether the terms in Eqn. (\ref{finalsc}) containing these
singularities cancel out. If we go back to Eqn. (\ref{gpcorr}) for the
correction to the grand potential, we see that the term ${\cal{D}}$
that goes as $q^{d-1}$ is inside a logarithm. Since $q^{d-1}$ occurs
as a product, we can simply take off the term $\log{(q^{d-1})}$, and
upon differentiating with respect to $\De$ should get zero. In other
words we naively expect no singular term in Eqn. (\ref{finalsc}). However
this is not quite correct since to prove that the denominator
${\cal{D}}$ vanished at $q=0$ we needed to replace $\la^{-1}$ using
the mean-field self-consistency equation, Eqn. (\ref{mfeqn}). We note that
although the two sides of Eqn. (\ref{mfeqn}) are numerically equal in
the mean-field case, their dependences on $\De$ differ -- $\la^{-1}$
gives zero under $\pa/\pa\De$, whilst $1/U$ does not. This discrepancy
then leads to the only singular term in Eqn. (\ref{finalsc}), which may be
written
\begin{equation}
\ln{\left({\De\over\De_0}\right)}_{mf}=
{1\over 4}\pi T\sum_\om {1\over\De^2 U^3}
\left[1-{\zeta\over U^3}\right]^{-1} 
T\sum_\Om\sum_q V_{\phi\phi}(q,\Om).
\end{equation}
Since this term tends to half the pair propagator contribution to
the suppression of $T_c$ when we let $\De\rightarrow 0$, we interpret
it as the phase fluctuation contribution. It is singular because of the
Mermin-Wagner-Hohenberg theorem\cite{MWH} 
which tells us that we cannot have broken
symmetry states in 2D systems at finite temperature. In the following we
will be mainly interested in the correction to $T_c$ due to Coulomb
interaction and so will ignore this term.

\section{First Order Correction to the Transition Temperature}

We can now evaluate the first order correction to the transition
temperature by linearizing the order parameter self-consistency
equation with respect to $\De$. The former can also be obtained
directly from the normal state by calculating the pair propagator
$L(q,\Om)$ to first order, and looking for the instability at
$q=\Om=0$. $L$ is given by
\begin{equation}
L^{-1}(q,\Om)=\lambda^{-1}+P(q,\Om),
\end{equation}
where $P(q,\Om)$ is the pair polarization bubble. The zeroth order
polarization bubble $P_0(q,\Om)$ is shown in Fig. (6b) and leads
to the mean-field result
\begin{eqnarray}
L_0^{-1}(q,\Om)&=&N(0)\left[\log{\left({T\over T_{c00}}\right)}
+\psi\left({1\over 2}+{1\over 2\pi T\tau_s}+
{Dq^2+|\Om|\over 4\pi T}\right)-\psi\left({1\over 2}\right)\right]
\nonumber\\
&=&N(0)\left[\log{\left({T\over T_{c0}}\right)}
+\psi\left({1\over 2}+{1\over 2\pi T\tau_s}+
{Dq^2+|\Om|\over 4\pi T}\right)
-\psi\left({1\over 2}+{1\over 2\pi T\tau_s}\right)\right],
\end{eqnarray}
where $T_{c00}$ is the BCS transition temperature (the mean field
value in the absence of magnetic impurities), and $T_{c0}$ is the
mean field value for the system with magnetic impurities. A
correction to the polarization operator $\delta P(0,0)$ will lead
to a change in the transition temperature, which is defined as the
temperature at which the denominator of $L$ becomes zero, given by
\begin{equation}
\log{\left({T_c\over T_{c0}}\right)} =
{\delta P(0,0)\over N(0)}.
\end{equation}
\par
If we look at Fig. (6) we see that there are 7 diagrams which
contribute to the first order correction to $T_c$. We will set
$\De=0$ in the order parameter result of Eqn. (\ref{finalsc})
to get the transition
temperature equation, and we will be able to identify the contribution
that comes from each of the $P_i$ diagrams. When we set
$\De\rightarrow 0$, then $\De u\rightarrow (|\om|+1/\tau_s)\hbox{sgn}(\om)$;
$\De U\rightarrow |\om|+1/\tau_s$; $V_{\rho\rho}\rightarrow 2V_C$;
$V_{\De\De}$ and $V_{\phi\phi}\rightarrow -L$;
$V_{\phi\rho}\rightarrow 2\Pi_{\phi\rho}LV_C$. The coherence factors
then become Heaviside functions that set the relative signs of the
frequencies
\begin{eqnarray}
1-{uu'\pm 1\over UU'} &\rightarrow&2\theta(-\om(\om+\Om))\nonumber\\
1+{uu'\pm 1\over UU'} &\rightarrow&2\theta(\om(\om+\Om)).
\end{eqnarray}
The two denominators, $D_{\pm}$, both become $Dq^2+|\Om|$ for $\om$, $\om+\Om$
of opposite sign; $Dq^2+|2\om+\Om|+2/\tau_s$ for  $\om$, $\om+\Om$
of the same sign. Making all these substitutions leads to
\begin{eqnarray}
P_1&=&-\pi N(0)T\sum_\om T\sum_\Om \sum_q \left[
{1\over (|\om|+1/\tau_s)^2}{1\over (Dq^2+|\Om|)}
+{2\over (|\om|+1/\tau_s)}{1\over (Dq^2+|\Om|)^2}\right]
V_C(q,\Om)\theta(-\om(\om+\Om))\nonumber\\
P_2&=&\pi N(0)T\sum_\om T\sum_\Om \sum_q
{1\over (|\om|+1/\tau_s)^2}{1\over Dq^2+|2\om+\Om|+2/\tau_s}
V_C(q,\Om)\theta(\om(\om+\Om))\nonumber\\
P_3&=&-\pi N(0)T\sum_\om T\sum_\Om \sum_q
{1\over (|\om|+1/\tau_s)(|\om+\Om|+1/\tau_s)}
\left[{1\over (Dq^2+|\Om|)}+{2/\tau_s\over(Dq^2+|\Om|)^2}\right]
V_C(q,\Om)\theta(-\om(\om+\Om))\nonumber\\
P_4&=&-\pi N(0)T\sum_\om T\sum_\Om \sum_q
{1\over (|\om|+1/\tau_s)(|\om+\Om|+1/\tau_s)}
{1\over (Dq^2+|2\om+\Om|+2/\tau_s)}
V_C(q,\Om)\theta(\om(\om+\Om))\nonumber\\
P_5&=&-\pi N(0)T\sum_\om T\sum_\Om \sum_q{1\over (|\om|+1/\tau_s)^2}\left[
{1\over (Dq^2+|2\om+\Om|+2/\tau_s)}+
{|\om|-1/\tau_s\over (Dq^2+|2\om+\Om|+2/\tau_s)^2}
\right]L(q,\Om)\theta(\om(\om+\Om))\nonumber\\
P_6&=&\pi N(0)T\sum_\om T\sum_\Om \sum_q
{1\over (|\om|+1/\tau_s)^2 (Dq^2+|\Om|)}
L(q,\Om)\theta(-\om(\om+\Om))\nonumber\\
P_7&=&4\pi^2 N(0)^2 T\sum_\Om \sum_q
\left[T\sum_\om {\hbox{sgn}(\om+\Om)\over (|\om|+1/\tau_s)
(Dq^2+|\om|+|\om+\Om|+2/\tau_s)\theta(\om(\om+\Om)))}\right]^2
V_C(q,\Om)L(q,\Om).
\end{eqnarray}
The assignment of terms to the polarization bubble diagram they
would arise from if we had done the calculation by that method is
unique, and can be summarised below:
\begin{itemize}
\item[] $P_1$ : term proportional to $V_C$, $\theta(-\om(\om+\Om))$
with no $|\om+\Om|+1/\tau_s$ denominator.
\item[] $P_2$ : term proportional to $V_C$, $\theta(\om(\om+\Om))$
with no $|\om+\Om|+1/\tau_s$ denominator.
\item[] $P_3$ : term proportional to $V_C$, $\theta(-\om(\om+\Om))$
with $|\om+\Om|+1/\tau_s$ denominator.
\item[] $P_4$ : term proportional to $V_C$, $\theta(\om(\om+\Om))$
with no $|\om+\Om|+1/\tau_s$ denominator.
\item[] $P_5$ : term proportional to $L$ and $\theta(-\om(\om+\Om))$.
\item[] $P_6$ : term proportional to $L$ and $\theta(\om(\om+\Om))$.
\item[] $P_7$ : term proportional to $LV_C$.
\end{itemize}

We find that these reduce to the results of I when we set
$1/\tau_s\rightarrow 0$, providing a useful consistency check on the 
present calculation.
\par
To evaluate the correction to $T_c$ we split it into two parts:
the Coulomb part consisting of those parts that contain a Coulomb
propagator, ($P_1-P_4$, $P_7$), and consequently require special
attention at $q=0$, and the fluctuation part consisting of those
terms that contain only a fluctuation propagator, ($P_5$, $P_6$).
Performing the $\om$-sum first we get for the Coulomb part
\begin{eqnarray}
\label{finaltc}
&&\log{\left({T_c\over T_{c0}}\right)}=
-T_c\sum_\Om \sum_q \left\{ {1\over 2\pi T_c}{Dq^2\over\Om^2-(Dq^2)^2}
\psi'\left({1\over 2}+{1\over 2\pi T_c\tau_s}+{|\Om|\over 2\pi T_c}
\right)\right.\nonumber\\
&+&\left.\left[{2Dq^2[\Om^2+(Dq^2)^2]\over |\Om|[\Om^2-(Dq^2)^2]^2}
-{\displaystyle\left({2\over\tau_s}\right)Dq^2\over 
\displaystyle|\Om|\left(|\Om|+{2\over\tau_s}\right)\left(|\Om|+Dq^2\right)}
\right]\left[\psi\left({1\over 2}+{1\over 2\pi T_c\tau_s}
+{|\Om|\over 2\pi T_c}\right)-\psi\left({1\over 2}+{1\over 2\pi T_c\tau_s}
\right)\right]\right.\nonumber\\
&-&\left.\di{4(Dq^2)^2\over [\Om^2-(Dq^2)^2]^2}
{\displaystyle\left[\psi\left({1\over 2}+{1\over 2\pi T_c\tau_s}
+{|\Om|\over 2\pi T_c}\right)
-\psi\left({1\over 2}+{1\over 2\pi T_c\tau_s}\right)\right]^2\over
\displaystyle\left[\psi\left({1\over 2}+{1\over 2\pi T_c\tau_s}
+{Dq^2+|\Om|)\over 4\pi T_c}\right)-
\psi\left({1\over 2}+{1\over 2\pi T_c\tau_s}\right)\right]}\right\}V_C(q,\Om).
\end{eqnarray}
Since the worst singularity possible in $V_C(q,\Om)$ at $q=0$ goes
as $1/q^2$, the overall $q^2$ factor multiplying $V_C$
in the above expression means that this singularity is removed.
It follows that the removal of the $q=0$ singularity in the Coulomb
part is unaffected by the addition of magnetic impurities -- this
is because it is a general feature enforced by gauge invariance, as
we will show in Appendix B.
\par
To calculate the Coulombic suppression term of Eqn. (\ref{finaltc}), 
we change variables to
$m=\Om/2\pi T$ and $y=Dq^2/2\pi T$, noting that $\sum_q=\int dy/(8\pi^2 DT)$,
and
\begin{equation}
N(0)V_C(q,\Om)=N(0)\left[{q^2\over 4\pi e^2}
+{2N(0)Dq^2\over Dq^2+|\Om|}\right]^{-1}
\approx {Dq^2+|\Om|\over 2Dq^2}={m+y\over 2y}.
\end{equation}
This leads to the result
\begin{eqnarray}
\label{finaltc1}
\log{\left({T_c\over T_{c0}}\right)}=
\displaystyle -{1\over 8\pi^2N(0)D}\sum_{m=1}^M\int_0^M dy&&
\left\{{1\over m-y}\psi'\left({1\over 2}+\alpha\right)\right.\nonumber\\
&+&\left[{2y(m^2+y^2)\over m(m^2-y^2)^2}
-{2\alpha y\over m(m+2\alpha)(m+y)}\right]
\left[\psi\left({1\over 2}+\alpha+m\right)-\psi\left({1\over 2}
+\alpha\right)\right]\nonumber\\
&-&\left.{4y\over (m-y)(m^2-y^2)}
{\left[\psi\left({1\over 2}+\alpha+m\right)-
\left({1\over 2}+\alpha\right)\right]^2\over
\left[\psi\left({1\over 2}+\alpha+{m+y\over 2}\right)-
\psi\left({1\over 2}+\alpha\right)\right]}\right\},
\end{eqnarray}
where $\alpha=1/2\pi T_c\tau_s$ and the upper cutoff $M=1/2\pi T_c\tau$.
The leading order term is that which goes like $1/y$ at large $y$, leading
to logarithmic behavior. To isolate this, we add and subtract the term
\begin{eqnarray}
&&\sum_{m=1}^{M}\int_0^M {dy\over (m+y)}
\left\{\left({2\over m}-{2\alpha\over m(m+2\alpha)}\right)
\left[\psi\left({1\over 2}+\alpha+m\right)
-\psi\left({1\over 2}+\alpha\right)\right]
-\psi'\left({1\over 2}+\alpha\right)\right\}\nonumber\\
=&&\sum_{m=1}^{M} \ln{\left({M+m\over m}\right)}
\left\{{2(m+\alpha)\over m(m+2\alpha)}
\left[\psi\left({1\over 2}+\alpha+m\right)
-\psi\left({1\over 2}+\alpha\right)\right]
-\psi'\left({1\over 2}+\alpha\right)\right\},
\end{eqnarray}
to give the result
\begin{eqnarray}
\label{onept}
\ln{\left({T_c\over T_{c0}}\right)}=&-&{R_{\square}\over R_0}\left\{
\sum_{m=1}^M \ln{\left({M+m\over m}\right)}
\left[{2(m+\alpha)\over m(m+2\alpha)} 
\left[\psi\left({1\over 2}+\alpha+m\right)
-\psi\left({1\over 2}+\alpha\right)\right]
-\psi'\left({1\over 2}+\alpha\right)\right]\right.\nonumber\\
&-&\sum_{m=1}^M\int_0^M dy{4y\over (m-y)(m^2-y^2)}
\left[{\left[\psi\left({1\over 2}+\alpha+m\right)-
\psi\left({1\over 2}+\alpha\right)\right]^2\over
\left[\psi\left({1\over 2}+\alpha+{m+y\over 2}\right)\right]}\right.\nonumber\\
&-&\left.\left.\left[\psi\left({1\over 2}+\alpha+m\right)-
\psi\left({1\over 2}+\alpha\right)\right]
+{y-m\over 2}\psi'\left({1\over 2}+\alpha\right)\right]\right\},
\end{eqnarray}
where we have noted that $1/8\pi^2N(0)D=R_{\square}/R_0$. 
We could now proceed to evaluate this expression, but before we do
so, let us consider the domain of validity of the first-order perturbative
result.

\section{Beyond Perturbation Theory}

Since we now have the full first-order perturbative correction
to the transition temperature due to the effect of disorder on
the Coulomb interaction, we could in principle plot curves of 
$T_c(R_{\square},1/\tau_s)$ and compare to experiment. 
However the curves of $T_c$ vs
$R_{\square}$ for different values of $1/\tau_s$ would simply
be exponential decays with different initial slopes. First-order
perturbation theory is unable to treat the strong disorder region,
and so cannot lead to the complete destruction of the superconductivity
by non-magnetic disorder.
If we are to consider the effects of arbitrary disorder strength, we
must work beyond perturbation theory. In what follows we discuss two
methods of doing this, and compare the results we obtain from them.

The simplest way to proceed is to ``self-consistently'' solve the
first-order perturbative expression of Eq. (\ref{onept}). This simply
means that we replace $T_{c0}$ by $T_c$ on the right-hand side of
Eq. (\ref{onept}), and solve the implicit equation we obtain for $T_c$
which has the form
\begin{equation}
\label{oneptsc}
\ln{\left({T_c\over T_{c0}}\right)}=\psi\left({1\over 2}\right)
-\psi\left({1\over 2}+{1\over 2\pi T_c\tau_s}\right)
-{R_{\square}\over R_0}f(T_c,1/\tau_s).
\end{equation}
Here $f(T_c,1/\tau_s)$ is the complicated expression on the right-hand side
of Eq. (\ref{onept}),
whilst the first term is just the mean-field suppression of $T_c$ by the
magnetic impurities. From our knowledge of the situation without magnetic
impurities, we know that unphysical re-entrance problems may arise with
the solution of this equation, and we should not take it too seriously
in the region where superconductivity is strongly suppressed.

The fact that we cannot trust the results obtained from this 
``self-consistent'' theory leads us to ask the
question of how to correctly go beyond first-order perturbation theory.
The best approach is to derive the effective field theory from which the
perturbation series may be deduced -- in this case a non-linear sigma
model\cite{Fink83} -- and treat this using the renormalization group. 
This has been done by Finkel'stein\cite{Fink87}
for the system without magnetic impurities, but
has the problem that it is very difficult, and would become even more
so if magnetic impurities were added. Recently Oreg and Finkel'stein\cite{OF}
demonstrated that the same results could be obtained using a much
simpler non-perturbative resummation technique, which we show
diagrammatically in Fig. (7). The method uses a featureless Coulomb
interaction of magnitude $N(0)V_C=1/2$, consistent with the cancellation
of the $1/q^{d-1}$ divergence discussed earlier, and keeps only diagrams
3 and 4 of Fig. (6), since they give the greatest contribution. This 
leads to the equation for the pair scattering amplitude, 
$\Gamma(\om_n,\om_l)$,
\begin{equation}
\label{of}
\Gamma(\om_n,\om_l)=-|\lambda|+t\Lambda(\om_n,\om_l)
-\pi T\sum_{m=-(M+1)}^M [-|\lambda|+t\Lambda(\om_n,\om_m)]
{1\over |\om_m|+1/\tau_s}
\Gamma(\om_m,\om_l),
\end{equation}
where $\om_n=2\pi T(n+1/2)$ is a Fermi Matsubara frequency, and the upper
cut-off $M=1/2\pi T\tau$. The amplitude $\Lambda(\om_n,\om_l)$ is given by 
\begin{equation}
\Lambda(\om_n,\om_l)=\cases{
\di\ln{\left[{1\over (|\om_n|+|\om_l|)\tau}\right]}&$\quad\om_n\om_l<0$\cr
\di\ln{\left[{1\over (|\om_n|+|\om_l|+2/\tau_s)\tau}\right]}
&$\quad\om_n\om_l>0$\cr}
\end{equation}
where the breaking of time-reversal invariance by the spin-flip scattering
means that $\Lambda$ has a different form depending upon the relative signs
of its two Matsubara frequencies. If we treat the $\Gamma(\om_n,\om_m)$ as
elements of a matrix $\hat{\Gamma}$, the matrix equation for $\hat{\Gamma}$
can be solved to yield
\begin{equation}
\hat{\Gamma}=\hat{\om}^{1/2}(\hat{I}-|\lambda|\hat{\Pi})^{-1}
\hat{\om}^{-1/2}(-|\lambda|\hat{1}+t\hat{\Lambda}),
\end{equation}
where
\begin{equation}
\hat{\Pi}={1\over 2}\hat{\om}^{-1/2}
[\hat{1}-|\lambda|^{-1}t\hat{\Lambda}]\hat{\om}^{-1/2},
\end{equation}
$\hat{\om}_{nm}=(n+1/2+\alpha)\delta_{nm}$,
$\hat{\Lambda}_{nm}=\Lambda(\om_n,\om_m)$, $\hat{1}_{nm}=1$ and
$\hat{I}_{nm}=\delta_{nm}$. The matrix $\hat{\Gamma}$ becomes singular when
an eigenvalue of $\hat{\Pi}$ equals $1/|\lambda|$, and this signals the onset
of superconductivity. Note that the matrix $\hat{\Pi}$ depends on temperature
both through the temperature dependence of its elements, and also through its
rank $2M$. To find $T_c$, we start at the BCS value $T_{c0}$, which corresponds
to a value of $M$ given by $M_0=1/2\pi T_{c0}\tau$. We decrease the 
temperature $T$ by increasing the upper cut-off $M$ successively by one.
For each value of $M$, we construct the matrix $\hat{\Pi}$, and diagonalise
it. When its lowest eigenvalue equals $1/|\lambda|$, we have reached the
transition temperature $T_c$, which is given by $T_c/T_{c0}=M_0/M$. This
method allows us to go to as low a temperature as we like, provided that
we are prepared to diagonalize large enough matrices.

We will now plot curves of $T_c$ vs $R_{\square}$ for fixed $1/\tau_s$,
and $T_c$ vs $1/\tau_s$ for fixed $R_{\square}$, derived both from the
self-consistent perturbation theory of Eqn. (\ref{oneptsc}), and from the
non-perturbative resummation approach of Eqn. (\ref{of}). This is done
in Fig.~(8), and we see that the two approaches are in rough agreement. 
The resummation technique is seen to remove the re-entrance problem which 
occurs in the $\alpha=0$ curve at large $R_{\square}$, but surprisingly
this re-entrance seems to be partially cured by the presence of magnetic
impurities.

The above curves are fine from the theorist's point of view, but 
experimentally what is measured is the suppression of $T_c$ by a
certain fixed amount of magnetic impurities as the thickness of the
superconductor is altered. The data is then presented in the form of
the pair-breaking per magnetic impurity which can be written as
\begin{equation}
\alpha'(R_{\square})={T_c(R_{\square},0)-T_c(R_{\square},1/\tau_s)\over
1/\tau_s},
\end{equation}
which we can also generate from our theoretical expressions.
The result will of course depend upon the magnitude of the value of
$\alpha$ we choose: we would like to choose $\alpha$ as small as possible
so that we are always in the linear regime of pair-breaking, but not too
small so that the difference is very sensitive to the discrete sums used
in the numerical calculation. A typical plot is shown in Fig. (9). If we
ignore the numerical noise we see that $\alpha'$ is roughly constant and
equal to its mean field value of $\pi^2/2$. It only appears to increase
as we approach the region where superconductivity is destroyed, and its
total variation is only about $10\%$ even if we include this region. This
is in agreement with the experimental data of Chervenak and Valles\cite{CV}.

\section{Discussion and Conclusions}

The main conclusion of this paper is that the effects of localization and
interaction do not lead to an appreciable change of pair-breaking rate per
magnetic impurity in disordered superconducting films provided that we are 
not too close to the 
superconductor-insulator transition. The experimental data agrees with
this theoretical prediction, and thus confirms the validity of the basic
model of $T_c$ suppression in disordered superconductors which consists of
the BCS interaction, Coulomb repulsion and static disorder. The fact that
the theoretical
prediction is obtained both from first-order perturbation theory, and from
a non-perturbative resummation technique, gives us increased confidence in
its validity. Our calculations demonstrate that the resummation technique is
a very powerful tool for going beyond perturbation theory which can be
adapted to a variety of situations. Moreover we find that the ad hoc
``self-consistent'' extension of first-order perturbation theory can give
sensible results even at values of $R_{\square}$ near the 
superconductor-insulator transition, at least in the presence of
pair-breaking. 

The effect of nonmagnetic disorder on pair-breaking in superconducting films
has previously been considered by Devereaux and Belitz\cite{Dev96} using a
model in which strong coupling effects are considered. Good agreement with
experiment is also obtained with this approach, although more fitting
parameters are required in this model. We note that only a single fitting
parameter -- the initial slope of the $T_c(R_{\square})$ curve -- is
required in our approach. Unfortunately we see that the experimental data
is unable to determine which, if any, of the two approaches is correct.
In support of our approach we note that it is a ``minimal model'' in the
sense that it contains the minimal physics to describe the system, and
requires the input of a single fitting parameter. However, this is not to
say that strong-coupling effects are not important in this system. 

Another important result which emerges from the approach based on the
grand-canonical potential is that the $1/q^2$ singularity of the disorder
screened Coulomb potential is always cancelled in first-order perturbation
theory. This removes the possibility of changing some experimental parameter
to obtain a strong suppression of $T_c$ from this singularity. We have shown
that this cancellation is enforced by gauge invariance, and leads us to
suspect that it occurs to all orders in perturbation theory. It is this
cancellation which makes it legitimate to use a featureless interaction in
the resummation technique.

\bigskip
\centerline {\bf ACKNOWLEDGEMENTS}
\medskip

We thank I. Aleiner, A.M. Finkel'stein and Y. Oreg for helpful discussions.
R.A.S. acknowledges support from the Nuffield Foundation.
V.A. is supported by the U.S. National Science Foundation under grant DMR-9805613.

\medskip
\appendix
\section{Calculation of Polarization Bubbles}
\medskip

In this appendix we give a detailed derivation of the polarization
bubbles, $\Pi_{ij}$, shown in Fig. (3). To evaluate these we must
first calculate the impurity ladder, $\Pi$, which is given by the
geometric series
\begin{equation}
\Pi=S+S\Ga_0 S+S\Ga_0 S\Ga_0 S+\dots,
\end{equation}
where
$\Ga_0$ is the impurity line
\begin{equation}
\Ga_0={1\over 2\pi N(0)\tau}\left[
\lambda_1\tau_3\sigma_0\otimes\tau_3\sigma_0+
\lambda_2 \left( \tau_0\sigma_1\otimes\tau_0\sigma_1+
\tau_0\sigma_2\otimes\tau_0\sigma_2+\tau_3\sigma_3\otimes\tau_3\sigma_3
\right)\right],
\end{equation}
and $1/\tau=1/\tau_0+1/\tau_s$ is the total impurity scattering
rate, $\lambda_1=\tau/\tau_0$, and $\lambda_2=\tau/3\tau_s$.
$S$ is the momentum sum of a direct product of Green functions
\begin{eqnarray}
S&=&\sum_k G(k,i\om)\otimes G(k+q,i\om+i\Om)\nonumber\\
&=&\displaystyle\pi N(0)\tau I\left[\tau_3\sigma_0\tau_3\sigma_0-
{(\zb-\Db\tau_1\sigma_3)(\zp-\Dp\tau_1\sigma_3)\over\ep\ep'}\right],
\end{eqnarray}
and $I$ is the integral
\begin{equation}
\label{Iint}
I={1\over\pi\tau}\int d\xi_k d\hat{\Om}
{\xi_k(\xi_k-{\bf q.v_F})\over (\xi_k^2-\ep^2)
[(\xi_k-{\bf q.v_F})^2-\ep'^2]}.
\end{equation}
Since we do not need the complete matrix structure of $\Pi$, but just
its traces with two matrices from the set $\ta$, $\tb$, $\tc$, we
actually evaluate the impurity dressed vertices $\Pi_j$ which have one
matrix from the above set inserted between the two terms of the direct
product in $\Pi$. These satisfy the equation
\begin{equation}
\label{pijeqn}
\Pi_j=S_j+S\Ga_0\Pi_j.
\end{equation}
Starting with $\Pi_{\De}$ we see that
\begin{eqnarray}
S_\De &=&\displaystyle 2\pi N(0)\tau I\left[\tc\ta\tc-
{(\zb+\Db\ta)\ta(\zp+\Dp\ta)\over\ep\ep'}\right]\nonumber\\
&=&\displaystyle 2\pi N(0)\tau I\left[-1-
{(\zb+\Db\ta)(\zp+\Dp\ta)\over\ep\ep'}\right]\ta\nonumber\\
&=&\displaystyle 2\pi N(0)\tau I(\apb-\bp\ta)\ta,
\end{eqnarray}
where the $\alpha$ and $\beta$ terms are coherence factors
\begin{equation}
\apm=1-{\zb\zp\pm\Db\Dp\over\ep\ep'}\quad;\quad
\apmb=\apm-2\quad;\quad
\bpm={\zp\Db\pm\zb\Dp\over\ep\ep'}.
\end{equation}
By inspection we see that $\Pi_{\De}$ must have the matrix form
\begin{equation}
\Pi_\De=2\pi N(0)\tau I[A+B\ta]\ta,
\end{equation}
and we now substitute this into Eqn. (\ref{pijeqn}) to deduce the
coefficients $A$ and $B$. To derive the second term on the RHS of
Eqn. (\ref{pijeqn}) we see that
\begin{equation}
\Ga_0\Pi_\De=-I\la_1(A-B\ta)\ta+3I\la_2(A+B\ta)\ta,
\end{equation}
and thus
\begin{eqnarray}
S\Ga_0\Pi_\De &=&\displaystyle2\pi N(0)\tau I\left\{\la_1(A+B\ta)\ta+
{(\zb+\Db\ta)\la_1(A+B\ta)\ta(\zp+\Dp\ta)\over\ep\ep'}\right.\nonumber\\
&&\left.\qquad\qquad\qquad\displaystyle -3\la_2(A-B\ta)\ta-
{(\zb+\Db\ta)\la_2(A-B\ta)\ta(\zp+\Dp\ta)\over\ep\ep'}\right\}\nonumber\\
&=&\displaystyle 2\pi N(0)\tau I\left\{(\la_1-3\la_2)A\left[1+
{(\zb+\Db\ta)(\zp+\Dp\ta)\over\ep\ep'}\right]\ta\right.\nonumber\\
&&\displaystyle\left.\qquad\qquad\qquad +(\la_1+3\la_2)B\ta\left[1-
{(\zb+\Db\ta)(\zp+\Dp\ta)\over\ep\ep'}\right]\ta\right\}\nonumber\\
&=&\displaystyle 2\pi N(0)\tau I[(\la_1-3\la_2)A(-\apb+\bp\ta)\ta
+(\la_1+3\la_2)B(\ap-\bp\ta)].
\end{eqnarray}
We can now equate the coefficients of $1$ and $\ta$ on the LHS and
RHS of Eqn. (\ref{pijeqn}) to obtain the linear equations for $A$ and $B$,
\begin{equation}
\left[\matrix{
1+I(\la_1-3\la_2)\apb&I(\la_1+3\la_2)\bp\cr
-I(\la_1-3\la_2)\bp&1-I(\la_1+3\la_2)\ap\cr}\right]
\left[\matrix{A\cr B\cr}\right]=
\left[\matrix{\apb\cr -\bp\cr}\right].
\end{equation}
This matrix equation can then be inverted by inverting the $2\times 2$
matrix and using the identity $\ap\apb=\bp^2$ to obtain
\begin{equation}
\left[\matrix{A\cr B\cr}\right]={1\over D_+}\left[\matrix{
1-I(\la_1+3\la_2)\ap& -I(\la_1+3\la_2)\bp\cr
I(\la_1-3\la_2)\bp&1+I(\la_1-3\la_2)\apb\cr}\right]
\left[\matrix{\apb\cr -\bp\cr}\right]
= {1\over D_+}\left[\matrix{\apb\cr -\bp\cr}\right],
\end{equation}
where the determinant $D_+$ can be written
\begin{eqnarray}
D_+&=&[1+I(\la_1-3\la_2)\apb][1-I(\la_1+3\la_2)\ap]
+I^2(\la_1+3\la_2)(\la_1-3\la_2)\bp^2\nonumber\\
&=& 1+I(\la_1-3\la_2)\apb-I(\la_1+3\la_2)\ap
=\displaystyle 1-2I\la_1+6I\la_2\left({\zb\zp+\Db\Dp\over\ep\ep'}\right).
\end{eqnarray}
Similar results are obtained for $\Pi_{\phi}$ and $\Pi_{\rho}$,
leading to the results
\begin{eqnarray}
\Pi_{\De} &=&2\pi N(0)\tau{I\over D_+}(\apb-\bp\ta)\ta\nonumber\\
\Pi_{\phi}&=&2\pi N(0)\tau{1\over D_-}(\amb-\bm\ta)\tb\nonumber\\
\Pi_{\rho}&=&2\pi N(0)\tau{I\over D_-}(\am-\bm\ta)\tc,
\end{eqnarray}
where
\begin{equation}
\label{denom}
D_{\pm} = 1-2I\la_1+6I\la_2\left({\zb\zp\pm\Db\Dp\over\ep\ep'}\right).
\end{equation}
If we evaluate the integral in Eqn. (\ref{Iint}) we find that $I$ is given by
\begin{equation}
\label{Ival}
2I\tau={1\over\Wb+\Wp}-{q^2v_F^2\over 2(\Wb+\Wp)^3},
\end{equation}
where
\begin{equation}
\Wb=\sqrt{\omb^2+\Db^2}\quad;\quad \Wb=\sqrt{\om'^2+\Dp^2}.
\end{equation}
From the second part of Eqn. (\ref{selfcons}) we see that we can write
\begin{equation}
\label{Wval}
\Wb+\Wp = {1\over\tau_0}-{1\over\tau_s}+\De U+\De U',
\end{equation}
and substituting Eqns. (\ref{Wval}) and (\ref{Ival})
into Eqn. (\ref{denom}) gives the result
\begin{equation}
\label{denomval}
D_{\pm}=\di 1-{\di\left[{1\over\tau_0}-{1\over\tau_s}
\left({uu'\mp 1\over UU'}\right)\right]
\over\di\left[{1\over\tau_0}-{1\over\tau_s}+\De U+\De U'\right]}
+Dq^2\tau\nonumber
\approx\left[Dq^2+\De U+\De U'+{1\over\tau_s}
\left({uu'\mp 1\over UU'}-1\right)\right]\tau.
\end{equation}
We can finally obtain the non-zero polarization bubbles $\Pi_{ij}$
by inserting the second matrix from the set $\ta$, $\tb$, $\tc$ into
$\Pi_j$ and taking the trace. This yields
\begin{eqnarray}
\label{pialla}
\Pi_{\De\De}(q,\Om) &=&\displaystyle\pi N(0)T\sum_\om \left[
{UU'+uu'-1\over UU'}\right]{1\over
\left(Dq^2+\De U+\De U'-\di{1\over\tau_s}\left[{UU'-uu'+1\over UU'}\right]
\right)}\nonumber\\
\Pi_{\phi\phi}(q,\Om) &=&\displaystyle \pi N(0)T\sum_\om \left[
{UU'+uu'+1\over UU'}\right]{1\over
\left(Dq^2+\De U+\De U'-\di{1\over\tau_s}\left[{UU'-uu'-1\over UU'}\right]
\right)}\nonumber\\
\Pi_{\rho\rho}(q,\Om) &=&\displaystyle-\pi N(0)T\sum_\om \left[
{UU'-uu'-1\over UU'}\right]{1\over
\left(Dq^2+\De U+\De U'-\di{1\over\tau_s}\left[{UU'-uu'-1\over UU'}\right]
\right)} + N(0)\nonumber\\
\Pi_{\phi\rho}(q,\Om) &=&\displaystyle-\pi N(0)T\sum_\om \left[
{u'-u\over UU'}\right]{1\over
\left(Dq^2+\De U+\De U'-\di{1\over\tau_s}\left[{UU'-uu'-1\over UU'}\right]
\right)}=-\Pi_{\rho\phi}(q,\Om).
\end{eqnarray}

\medskip
\section{Low-Momentum Singularities in Density and Phase Propagators}
\medskip

The identities
\begin{equation}
\label{finalident}
-\lambda^{-1}+\Pi_{\phi\phi}(0,\Om) = x\Pi_{\phi\rho}(0,\Om)~,~~~~
\Pi_{\phi\rho}(0,\Om) = -x\Pi_{\rho\rho}(0,\Om),
\end{equation}
where $x=\Om/2\De$, play a central role in the present paper and in I.  As
we have seen, their consequence
\begin{equation}
[-\lambda^{-1}+\Pi_{\phi\phi}(0,\Om)]\Pi_{\rho\rho}(0,\Om)
+\Pi_{\phi\rho}(0,\Om)^2 = 0
\end{equation}
leads to the potentials
$V_{\phi\phi}$, $V_{\phi\rho}$, and $V_{\rho\rho}$ having $1/q^{d-1}$
singularities at low momentum $q$ for all temperatures $0\le T\le T_c$
and all non-zero frequencies $\Om\ne 0$. The importance of
these identities suggests that they embody an underlying invariance
principle.  In this appendix we show that they are Ward identities
connected to charge conservation, which is very reasonable since the
impossibility of instantaneously moving the conserved screening charge
a finite distance is at the root of these singularites,  The ideas
at work here go back to Nambu's 1960 paper and its 
elaborations\cite{Namb60,Amb61,Schr}
and they are only included here for completeness. It is unfortunate
that the physical basis for
these identities was left obscure in I.

To avoid irrelevant notational complications, we shall work within the
$2\times 2$ Nambu space.  The $4\times 4$ space needed to deal with spin
flip scattering does not affect the general argument, and we shall in any
case explicitly verify the identities for this case later in this
appendix.

The `proper polarization parts', $\Pi$, are calculated within
a mean field approximation, in which the interactions are replaced
according to
\begin{equation}
\label{vmf}
V\rightarrow V_{MF} = \De Tr{\Psi^\dagger \tau_1 \Psi},
\end{equation}
which implies a choice of phase for the order parameter.  It is known that
the quasiparticles obtained in this approximation do not conserve charge,
because $V_{MF}$ does not commute with the electron density.  Since
the only other non-commuting part of the Hamiltonian is the kinetic energy,
the operator equation of motion for the density $\rho\equiv
Tr{\Psi^\dagger\tau_3\Psi}$ is
\begin{equation}
\label{eqmot1}
{\pa\rho\over\pa t} + \nabla\cdot\vec j = i [V_{MF},\rho] =
2\De Tr{\Psi^\dagger\tau_2\Psi},
\end{equation}
where $\vec j = Tr [\Psi^\dagger \vec\nabla \Psi - \Psi\vec\nabla\Psi^\dagger]
$ is the current density operator.
Eq. (\ref{eqmot1}) leads to the identity
\begin{eqnarray}
\label{eqmot2}
&~~~~~~~~~~~~{\pa\over\pa t_2} T \langle\Psi_i (\vec x_1,t_1) \rho
(\vec x _2, t_2) \Psi^\dagger_j (\vec x_3,t_3)\rangle = \nonumber\\
&\de (\vec x _1 -
\vec x_2) \de (t_1 - t_2) i [\tau _3 G(\vec x_2,t_2, \vec x_3,
t_3)]_{ij} -\de (\vec x_2 -\vec x_3) \de (t_2 - t_3) i [G(\vec
x_1,t_1, \vec x_2,t_2)\tau_3 ]_{ij}\nonumber\\ 
&-T \langle\Psi_i (\vec x_1,t_1)
\nabla_2\cdot\vec j (\vec x_2, t_2) \Psi^\dagger_j (\vec x_3,t_3)\rangle 
+ T\langle\Psi_i (\vec x_1,t_1) \rho_\phi
(\vec x_2, t_2) \Psi^\dagger_j (\vec x_3,t_3)\rangle,
\end{eqnarray}
where $T$ is the time ordering operator and we have defined $\rho_\phi\equiv
Tr{\Psi^\dagger\tau_2\Psi}$.  [The first two terms on the right of Eq.
(\ref{eqmot2})
come from the derivative of the time ordering operator.]
Multiplying Eq. (\ref{eqmot2}) by $\tau_3$, taking the trace, and Fourier
transforming in space and time leads in the limit of zero wave vector to
the identity
\begin{equation}
\Om~\Pi_{\rho \rho}(0, \Om) = 2 \De~\Pi_{\rho \phi} (0, \Om).
\end{equation}
On the other hand, multiplying by $\tau_2$ and performing these
same operations yields
\begin{eqnarray}
\Om~\Pi_{\phi \rho}(0,\Om)&=& - {i\over\beta} Tr [\tau_1 G] + 2\De
~\Pi_{\phi \phi}\nonumber\\ 
&=& 2 \De [-\lambda^{-1} + \Pi_{\phi \phi}].
\end{eqnarray}
In the second line above the self consistency equation within the mean
field approximation has been used.  To obtain Eq. (\ref{finalident}) 
we must note
that $\Pi_{\rho \phi}$ is antisymmetric in its indices
because of time reversal invariance---under which $\rho$ is symmetric and
$\rho_\phi$ antisymmetric.

Since the impurity interaction commutes with the charge density these
identities survive in any `conserving' approximation\cite{Kad61}
and, in particular, in our sum of all non-overlapping graphs.
In the main body of this paper, the mean field approximation was used as
a way station on the road to the single loop approximation of Section III.
There the phase of the order parameter is not fixed as in Eq.
(\ref{vmf}) but determined self consistently, which restores charge
conservation.\cite{PWA58,Rick58}

Finally, we shall verify explicitly that the identities
(\ref{finalident}) are satisfied by our calculated expressions.
We start with the equations for $-\lambda^{-1}+\Pi_{\phi\phi}$, $\Pi_{\phi\rho}$,
and $\Pi_{\rho\rho}$,
\begin{eqnarray}
\label{piforms}
-\lambda^{-1}+\Pi_{\phi\phi}(0,\Om) &=&  \di T\sum_{\om} \left[
{UU'+uu'+1\over UU'\left(U+U'-\zeta\left[{UU'-uu'-1\over UU'}\right]
\right)} -{1\over U}\right]\nonumber\\
\Pi_{\phi\rho}(0,\Om) &=&\di T\sum_{\om}
{u'-u\over UU'\left(U+U'-\zeta\left[{UU'-uu'-1\over UU'}\right]\right)}
\nonumber\\
\Pi_{\rho\rho}(0,\Om) &=&\di {1\over\pi} - T\sum_{\om}
{UU'-uu'-1\over UU'\left(U+U'-\zeta\left[{UU'-uu'-1\over UU'}\right]
\right)},
\end{eqnarray}
where we have removed the common factor
$\pi N(0)$ from each $\Pi$, and set $\De=1$ for algebraic convenience.
These factors can, of course, be replaced when we have finished.
\par
We will first prove the relationship between $-\lambda^{-1}+\Pi_{\phi\phi}$ 
and $\Pi_{\phi\rho}$, namely
\begin{equation}
\label{piident1}
-\lambda^{-1}+\Pi_{\phi\phi}(0,\Om) = {\Om\over 2}\Phi_{\phi\rho}(0,\Om).
\end{equation}
To proceed note that we can write
\begin{equation}
uu' = {1\over 2}[u^2+u'^2-(u'-u)^2]
={1\over 2}[U^2+U'^2-2-(u'-u)^2],
\end{equation}
from which it follows that
\begin{eqnarray}
\label{ident1}
UU'+uu'+1 &=& {1\over 2}[(U+U')^2-(u'-u)^2]\nonumber\\
UU'-uu'-1 &=& -{1\over 2}[(U'-U)^2-(u'-u)^2].
\end{eqnarray}
In the last term on the RHS of Eq. (\ref{piforms}) for $\Pi_{\phi\phi}$,
we can use the transformation $\om\leftrightarrow -\om'$, under which
the sum over $\om$ is invariant. This leads
to $u\leftrightarrow -u'$ and $U\leftrightarrow U'$, so that
\begin{equation}
\label{intermed1}
2T\sum_{\om} {1\over U} = T\sum_{\om} \left[ {1\over U}
+{1\over U'}\right] = T\sum_{\om} {U+U'\over UU'}.
\end{equation}
We can now write, using Eqs. (\ref{piforms}), (\ref{ident1}) and
(\ref{intermed1}),
\begin{eqnarray}
\label{intermed2}
2(-\lambda^{-1}+\Pi_{\phi\phi})-\Om\Pi_{\phi\rho}&=&
T\sum_{\om} \left[{(U+U')^2-(u'-u)^2+\Om(u'-u)\over UU'
\left(U+U'-\zeta\left[{UU'-uu'-1\over UU'}\right]\right)}
-{U+U'\over UU'}\right]\nonumber\\ &=&
T\sum_{\om}{(U+U')^2-(u'-u)(u'-u-\Om)-(U+U')
\left(U+U'-\zeta\left[{UU'-uu'-1\over UU'}\right]\right)\over
UU'\left(U+U'-\zeta\left[{UU'-uu'-1\over UU'}\right]\right)}.
\end{eqnarray}
From the definition of $u$ and $u'$ in Eqn. (\ref{transc}) we obtain the 
identity
\begin{eqnarray}
\label{ident2}
u'-u-\Om&=&(u'-\om')-(u-\om)=\zeta\left({u'\over U'}-{u\over U}\right)
\nonumber\\
\Rightarrow\quad
(u'-u)(u'-u-\Om)&=&\zeta(u'-u)\left({u'\over U'}-{u\over U}\right)
\nonumber\\
&=&\zeta\left[{u'^2\over U'}+{u^2\over U}-{uu'\over U'}-{uu'\over U}\right]
\nonumber\\
&=&\zeta\left[U+U'-{1\over U}-{1\over U'}-{uu'\over U}-{uu'\over U'}\right]
\nonumber\\
&=&\zeta(U+U')\left[{UU'-uu'-1\over UU'}\right].
\end{eqnarray}
It follows that the numerator in Eqn. (\ref{intermed2}) is zero,
and hence we have proved the required result (\ref{piident1}).
\par
We next prove the relation between $\Pi_{\rho\rho}$ and
$\Pi_{\phi\rho}$, namely
\begin{equation}
\label{piident2}
\Pi_{\phi\rho}(0,\Om)=-{\Om\over 2}\Pi_{\rho\rho}(0,\Om).
\end{equation}
We start by considering the sum
\begin{equation}
\lim_{\om_0\rightarrow\infty} T\sum_{\om=-(\om_0+\Om)}^{\om_0}
{u\over\Om U}.
\end{equation}
As $|\om|\rightarrow\infty$, $u\rightarrow (|\om|+\zeta)\hbox{sgn}(\om)$
and $u/U\rightarrow \hbox{sgn}(\om)$. Since there are $\Om/2\pi T$
more negative terms than positive, the sum becomes
\begin{equation}
\lim_{\om_0\rightarrow\infty} T\sum_{\om=-(\om_0+\Om)}^{\om_0}
{u\over\Om U} = -\left({\Om\over 2\pi T}\right)\left({T\over\Om}\right)
=-{1\over 2\pi}.
\end{equation}
We have chosen the limits so that we are able to make the usual
$\om\leftrightarrow -\om'$ transformation in this sum. It follows that
\begin{equation}
{1\over\pi} = -2T\sum_{\om} {u\over U\Om}
             =T\sum_{\om}{1\over\Om}\left({u'\over U'}
             -{u\over U}\right)
            = T\sum_{\om} {U+U'\over \Om(u'-u)}
              \left({UU'-uu'-1\over UU'}\right),
\end{equation}
where we first make the $\om\leftrightarrow -\om'$
transformation, and then use
(\ref{ident2}). We can then rewrite Eq. (\ref{piforms}) for $\Pi_{\rho\rho}$
in the form
\begin{eqnarray}
\label{intermed3}
\Pi_{\rho\rho} &=& T\sum_{\om}\left({UU'-uu'-1\over UU'}\right)
\left[{U+U'\over\Om(u'-u)}-{1\over
\left(U+U'-\zeta\left[{UU'-uu'-1\over UU'}\right]\right)}\right]\nonumber\\
&=& {T\over\Om}\sum_{\om}\left({UU'-uu'-1\over UU'}\right)
{(U+U')\left(U+U'-\zeta\left[{UU'-uu'-1\over UU'}\right]\right)
-(u-u')\Om\over (u'-u)
\left(U+U'-\zeta\left[{UU'-uu'-1\over UU'}\right]\right)}.
\end{eqnarray}
From the identity (\ref{ident2}) we see that the 
numerator of (\ref{intermed3}) can be rewritten as
\begin{eqnarray}
&&(U+U')\left(U+U'-\zeta\left[{UU'-uu'-1\over UU'}\right]\right)
-(u'-u)\Om\nonumber\\
&=& (U+U')^2-(u'-u)(u'-u-\Om)-(u'-u)\Om\nonumber\\
&=& (U+U')^2-(u'-u)^2,
\end{eqnarray}
and inserting the identity (\ref{ident2}) into the first factor in
(\ref{intermed3}), we get
\begin{equation}
\Pi_{\rho\rho}= -{T\over 2\Om}\sum_{\om}
{[(U'-U)^2-(u'-u)^2][(U+U')^2-(u'-u)^2]\over (u'-u)UU'
\left(U+U'-\zeta\left[{UU'-uu'-1\over UU'}\right]\right)}.
\end{equation}
Multiplying out the numerator yields
\begin{eqnarray}
&&[(U'-U)^2-(u'-u)^2][(U+U')^2-(u'-u)^2]\nonumber\\
&=& (u'^2-u^2)^2-2(u'-u)^2(U^2+U'^2)+(u'-u)^4\nonumber\\
&=& (u'-u)^2[(u+u')^2-2u^2-2u'^2-4+(u'-u)^2]\nonumber\\
&=& -4(u'-u)^2,
\end{eqnarray}
from which it follows that
\begin{equation}
\Pi_{\rho\rho}(0,\Om)=
-{2T\over\Om}\sum_{\om}{u'-u\over UU'
\left(U+U'-\zeta\left[{UU'-uu'-1\over UU'}\right]\right)}
=-{2\over\Om}\Pi_{\phi\rho}(0,\Om),
\end{equation}
completing our proof of the result (\ref{piident2}).

\medskip
\section{Evaluating Derivatives of $\Pi_{ij}$ with respect to $\De$}
\medskip

In this appendix we evaluate the derivatives of the polarization bubbles
$\Pi_{ij}$ with respect to the order parameter $\De$ so that we may
evaluate the first order correction to the order parameter
self-consistency equation. The formulas for the $\Pi_{ij}$ are given in
Eqn. (\ref{pialla}), and we see that the derivative can operate either on the
coherence factor or the denominator present in these expressions.
\par
The difficulty in evaluating these derivatives arises because $u(\om)$
satisfies the transcendental equation
\begin{equation}
{\om\over\De}=u\left[1-{1\over\De\tau_s}{1\over (u^2+1)^{1/2}}\right],
\end{equation}
from which it follows that
\begin{equation}
-{\om\over\De^2}={\pa u\over\pa\De}\left[1-{1\over\De\tau_s}
{1\over (u^2+1)^{3/2}}\right]+{1\over\De^2\tau_s}{u\over (u^2+1)^{1/2}},
\end{equation}
and thus
\begin{equation}
{\pa u\over\pa\De}=-{u\over\De}\left[1-{1\over\De\tau_s}
{1\over (u^2+1)^{3/2}}\right]^{-1},
\end{equation}
with a similar result for $u'$.
\par
We first consider the effect of $\pa/\pa\De$ on the coherence factors
present in the $\Pi_{ij}$. We see that it suffices to evaluate the
derivatives
\begin{equation}
\label{diffs}
{\pa\over\pa\De}\left\{ {uu'\over UU'}\quad;\quad {1\over UU'}
\quad;\quad {u'-u\over UU'} \right\}.
\end{equation}
The first term in Eqn. (\ref{diffs}) gives
\begin{equation}
\label{diff1}
{\pa\over\pa\De}\left[{uu'\over UU'}\right]=
{u'\over U'}\left[{1\over U}-{u^2\over U^3}\right]{\pa u\over\pa\De}
+(u\leftrightarrow u')
= -{1\over\De}{uu'\over U^3U'}\left[1-{\zeta\over U^3}\right]^{-1}
+(u\leftrightarrow u').
\end{equation}
Since this expression will occur inside a sum over $\om$, and will
multiply an expression that is invariant under the transformation
$\om\leftrightarrow -\om'$, we see that the two terms in Eqn. (\ref{diff1})
will give equal results. Thus
\begin{equation}
{\pa\over\pa\De}\left[{uu'\over UU'}\right]\equiv
-{2\over\De}{uu'\over U^3U'}\left[1-{\zeta\over U^3}\right]^{-1}.
\end{equation}
The second term in Eqn. (\ref{diffs}) gives
\begin{equation}
{\pa\over\pa\De}\left[{1\over UU'}\right] = -{1\over U'}{u\over U^3}
{\pa u\over\De}+(u\leftrightarrow u')
\equiv {2\over\De}{u^2\over UU'}\left[1-{\zeta\over U^3}\right]^{-1},
\end{equation}
whilst the third term in Eqn. (\ref{diffs}) gives
\begin{equation}
{\pa\over\pa\De}\left[u'-u\over UU'\right] =
-{u'\over U'}{u\over U^3}{\pa u\over\pa\De} -
{1\over U'}{1\over U^3}{\pa u\over\pa\De}+(u\leftrightarrow u')
\equiv {2\over\De}{u(uu'+1)\over U^3U'}
\left[1-{\zeta\over U^3}\right]^{-1}.
\end{equation}
The effect of $\pa/\pa\De$ on the coherence factors can then be
summarised in the form
\begin{eqnarray}
\label{coheqn}
{\pa\over\pa\De}\left[{uu'\pm 1\over UU'}\right]
&\equiv& -{2\over\De}{u(u'\mp u)\over U^3U'}
\left[1-{\zeta\over U^3}\right]^{-1}\nonumber\\
{\pa\over\pa\De}\left[{u'-u\over UU'}\right]
&\equiv& {2\over\De}{u(uu'+1)\over U^3U'}
\left[1-{\zeta\over U^3}\right]^{-1}.
\end{eqnarray}
\par
Next we must consider the effect of $\pa/\pa\De$ on the two
denominators
\begin{equation}
D_{\pm}=Dq^2+\De U+\De U'-{1\over\tau_s}
\left(1-{uu'\mp 1\over UU'}\right).
\end{equation}
The last term on the RHS is a coherence factor, and its derivative can
be read off from Eqn. (\ref{coheqn}) above. The only terms then to consider
are $\De U$ and $\De U'$, which, of course, will give identical results
after summation over $\om$. We see that
\begin{equation}
{\pa\over\pa\De}(\De U) = U + \De {u\over U^3}{\pa u\over\pa\De}
= U - {u^2\over U^3}\left[1-{\zeta\over U^3}\right]^{-1}
= {1\over U}\left[1-{\zeta\over U}\right]
\left[1-{\zeta\over U^3}\right]^{-1}.
\end{equation}
From this we obtain the final result
\begin{equation}
{\pa\over\pa\De}D_{\pm} = \left\{ {1\over U}-{\zeta\over U^2}
\left(1+{u(u'-u)\over UU'}\right)\right\}
\left[1-{\zeta\over U^3}\right]^{-1}.
\end{equation}
\par
Having now evaluated the action of $\pa/\pa\De$ on all the components of
the polarization bubbles, $\Pi_{ij}$, we can now write down the results for
the $\pa\Pi_{ij}/\pa\De$,
\begin{eqnarray}
{\pa\Pi_{\De\De}\over\pa\De} &=& -2\pi N(0)T\sum_\om \left\{
\left[1-{\zeta\over U^3}\right]^{-1}\times
{1\over\De}{u(u'+u)\over U^3U'}{1\over D_{+}}+
\left(1+{uu'-1\over UU'}\right)\left\{{1\over U}-{\zeta\over U^2}
\left(1+{u(u'-u)\over UU'}\right)\right\}{1\over D_{+}^2}\right\}\nonumber\\
{\pa\Pi_{\phi\phi}\over\pa\De} &=& -2\pi N(0)T\sum_\om \left\{
\left[1-{\zeta\over U^3}\right]^{-1}\times
{1\over\De}{u(u'-u)\over U^3U'}{1\over D_{-}}+
\left(1+{uu'+1\over UU'}\right)\left\{{1\over U}-{\zeta\over U^2}
\left(1+{u(u'+u)\over UU'}\right)\right\}{1\over D_{-}^2}\right\}\nonumber\\
{\pa\Pi_{\rho\rho}\over\pa\De} &=& -2\pi N(0)T\sum_\om \left\{
\left[1-{\zeta\over U^3}\right]^{-1}\times
{1\over\De}{u(u'-u)\over U^3U'}{1\over D_{-}}-
\left(1-{uu'+1\over UU'}\right)\left\{{1\over U}-{\zeta\over U^2}
\left(1+{u(u'+u)\over UU'}\right)\right\}{1\over D_{-}^2}\right\}\nonumber\\
{\pa\Pi_{\phi\rho}\over\pa\De} &=& -2\pi N(0)T\sum_\om \left\{
\left[1-{\zeta\over U^3}\right]^{-1}\times
{1\over\De}{u(uu'+1)\over U^3U'}{1\over D_{-}}-
\left({u'-u\over UU'}\right)\left\{{1\over U}-{\zeta\over U^2}
\left(1+{u(u'+u)\over UU'}\right)\right\}{1\over D_{-}^2}\right\}.
\end{eqnarray}

\newpage

\centerline{\bf FIGURES}

\begin{figure}
\centerline{\psfig{figure=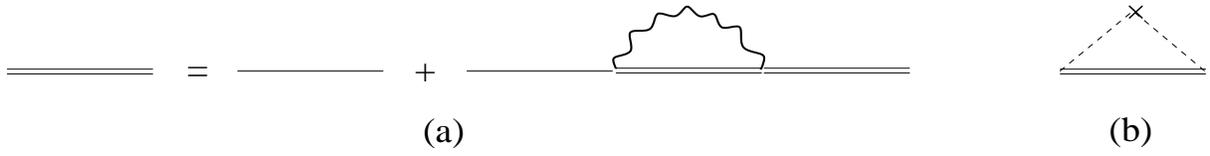,width=16cm}}
\medskip
\caption{The electron Green function in a superconductor.
(a) Self-energy for a clean superconductor. The wiggly line is
the BCS interaction. (b) Extra self-energy diagram needed for
dirty superconductor.}
\end{figure}

\newpage

\begin{figure}
\centerline{\psfig{figure=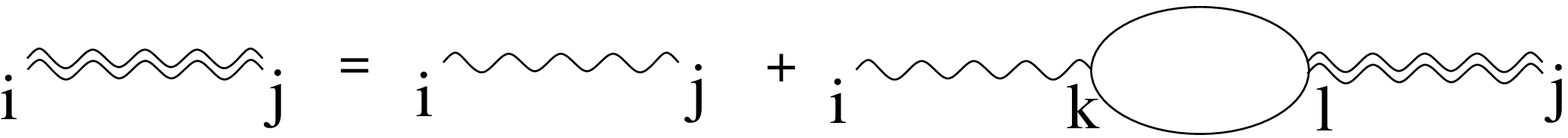,width=14cm}}
\medskip
\caption{Definition of the screened potential $V_{ij}$
in terms of the polarization bubble $\Pi_{ij}$ and bare
potential $V^0_{ij}$.}
\end{figure}

\newpage

\begin{figure}
\centerline{\psfig{figure=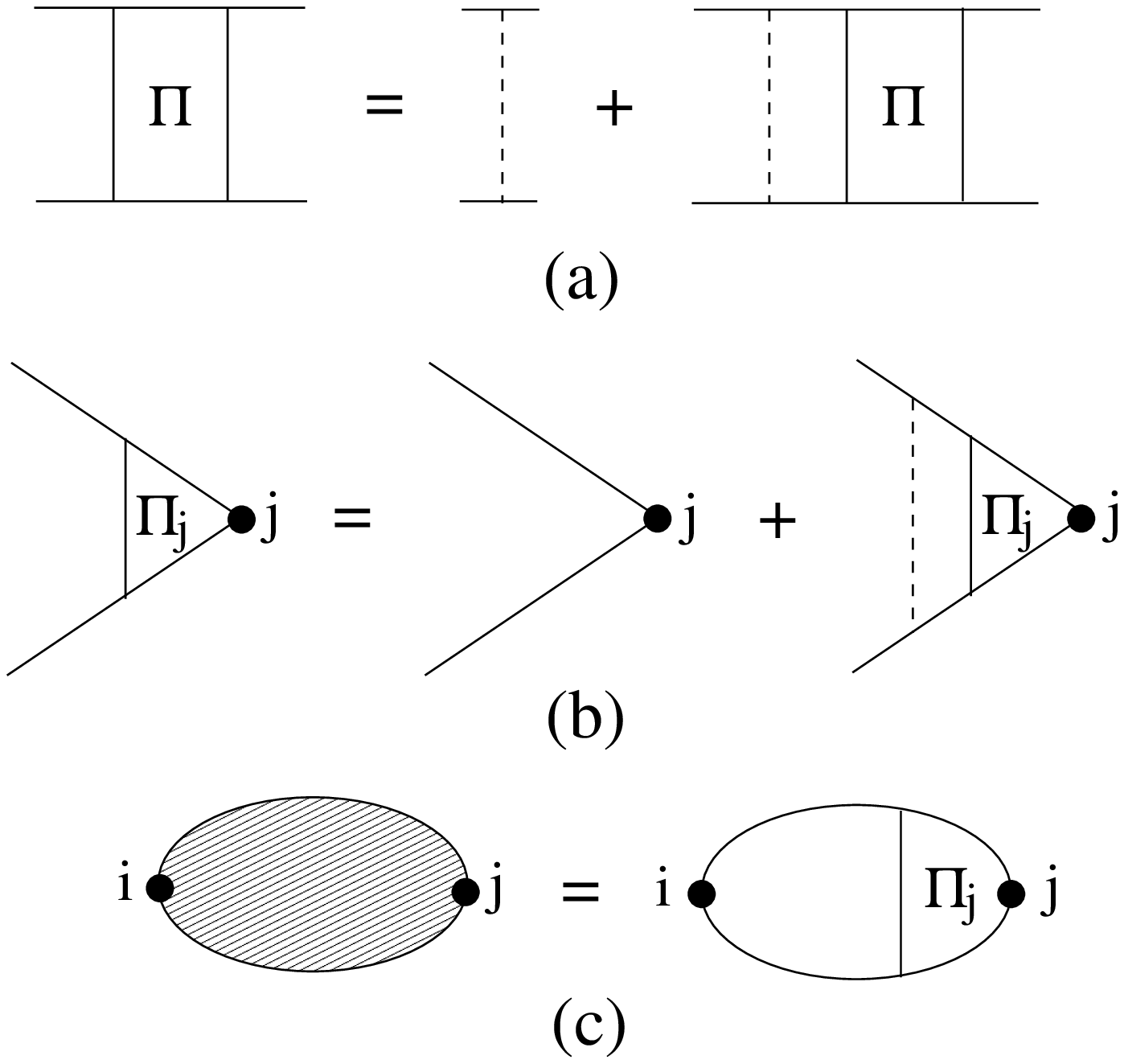,width=8cm}}
\medskip
\caption{Definition of the polarization bubbles $\Pi_{ij}$.
(a) The geometric series for the ladder $\Pi$.
(b) the geometric series for the vertex function $\Pi_j$ which
is obtained from $\Pi$ by taking the trace at one end with a Pauli
matrix.
(c) The polarization bubble $\Pi_{ij}$ is obtained from the vertex
operator $\Pi_j$ by taking the trace with a Pauli matrix at the open
end.}
\end{figure}

\newpage

\begin{figure}
\centerline{\psfig{figure=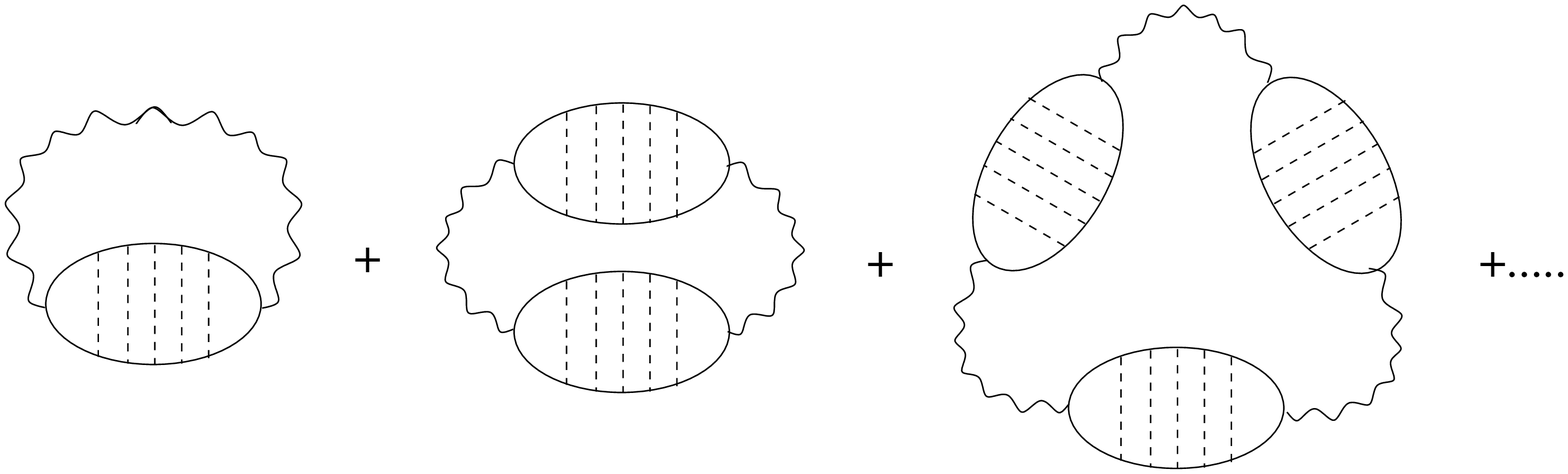,width=12cm}}
\medskip
\caption{The first-order correction to the grand canonical potential.
This has the form of a ``string of bubbles'', where the wiggly lines
can be either the bare Coulomb or BCS interaction, and the bubbles are
any of the non-zero polarization bubbles.}
\end{figure}

\newpage

\begin{figure}
\centerline{\psfig{figure=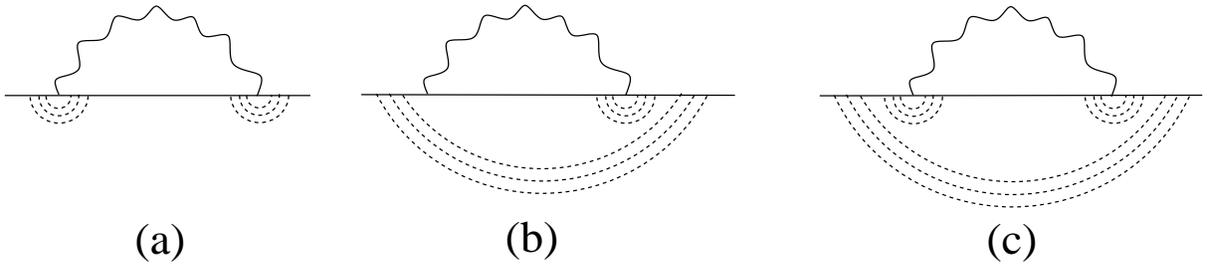,width=16cm}}
\medskip
\caption{The equivalent Eliashberg-like self-energy diagrams for the
correction to the order parameter $\De$. (a) The two-ladder diagrams
are obtained by differentiating the diffusion propagator term in the
$\Pi_{ij}$ with respect to $\De$. (b) The one-ladder diagrams are obtained
by differentiating the coherence factor term in $\Pi_{ij}$ with respect
to $\De$. (c) No three-ladder terms are obtained by differentiating
$\Om_1(\De)$ with respect to $\De$, and direct calculation of these
diagrams shows that they equal zero.}
\end{figure}

\newpage

\begin{figure}
\centerline{\psfig{figure=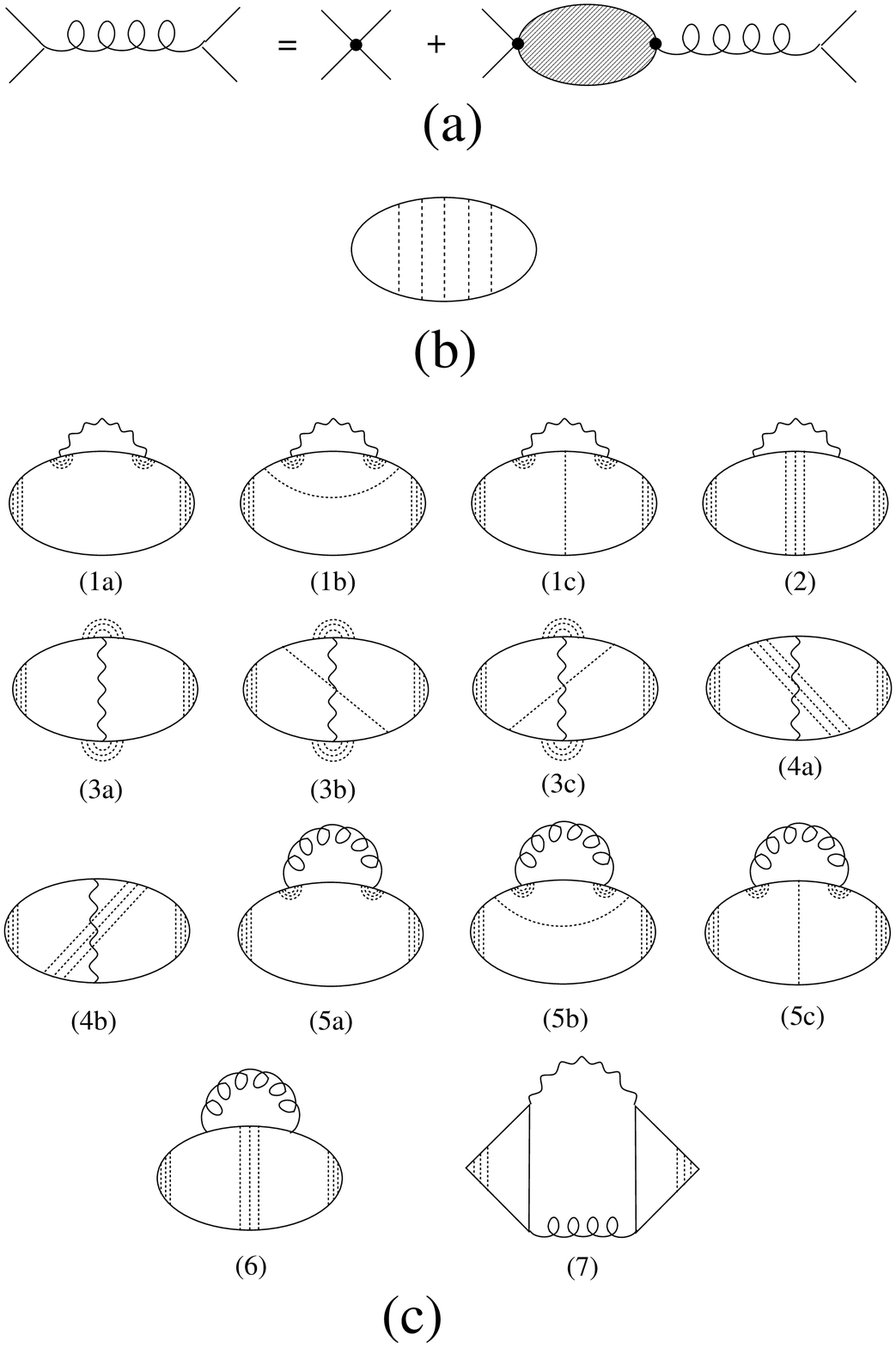,width=12cm}}
\medskip
\caption{The first-order correction to the pair propagator. The value
of temperature at which this first diverges is the transition temperature
$T_c$.
(a) Definition of pair propagator in terms of BCS interaction and
pair polarization bubble.
(b) Zeroth-order (mean field) pair polarization bubble.
(c) The 7 diagrams which contribute to the first-order correction to
the pair polarization bubble. The wiggly line is the screened Coulomb
interaction, whilst the spring-like line is the pair propagator, as defined
in (a).}
\end{figure}

\newpage

\begin{figure}
\centerline{\psfig{figure=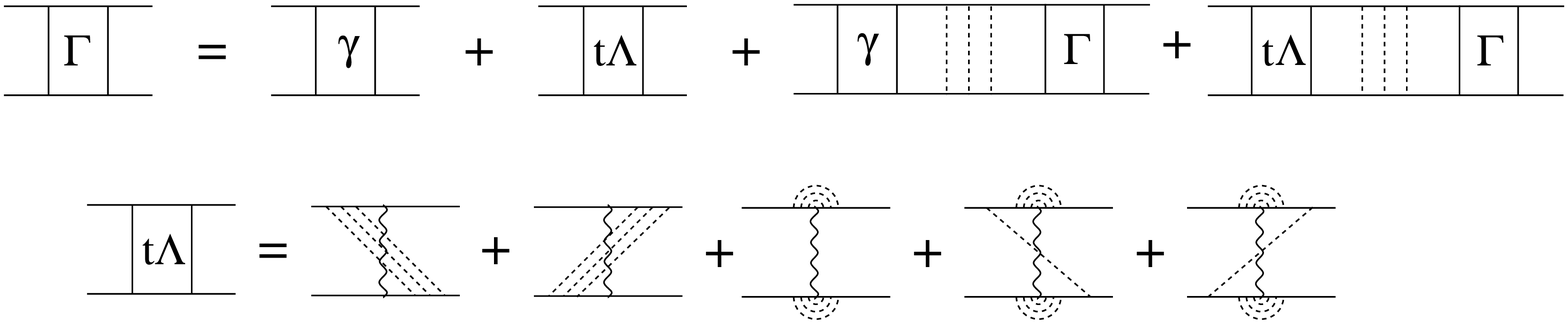,width=16cm}}
\medskip
\caption{Diagrammatic equation for the scattering amplitude matrix
$\Gamma(\om_n,\om_l)$. Block $\gamma$ is the BCS interaction.
Block $t\Lambda$ is the correction to the effective interaction
caused by the interplay of Coulomb interaction and disorder.}
\end{figure}

\newpage

\begin{figure}
\centerline{\psfig{figure=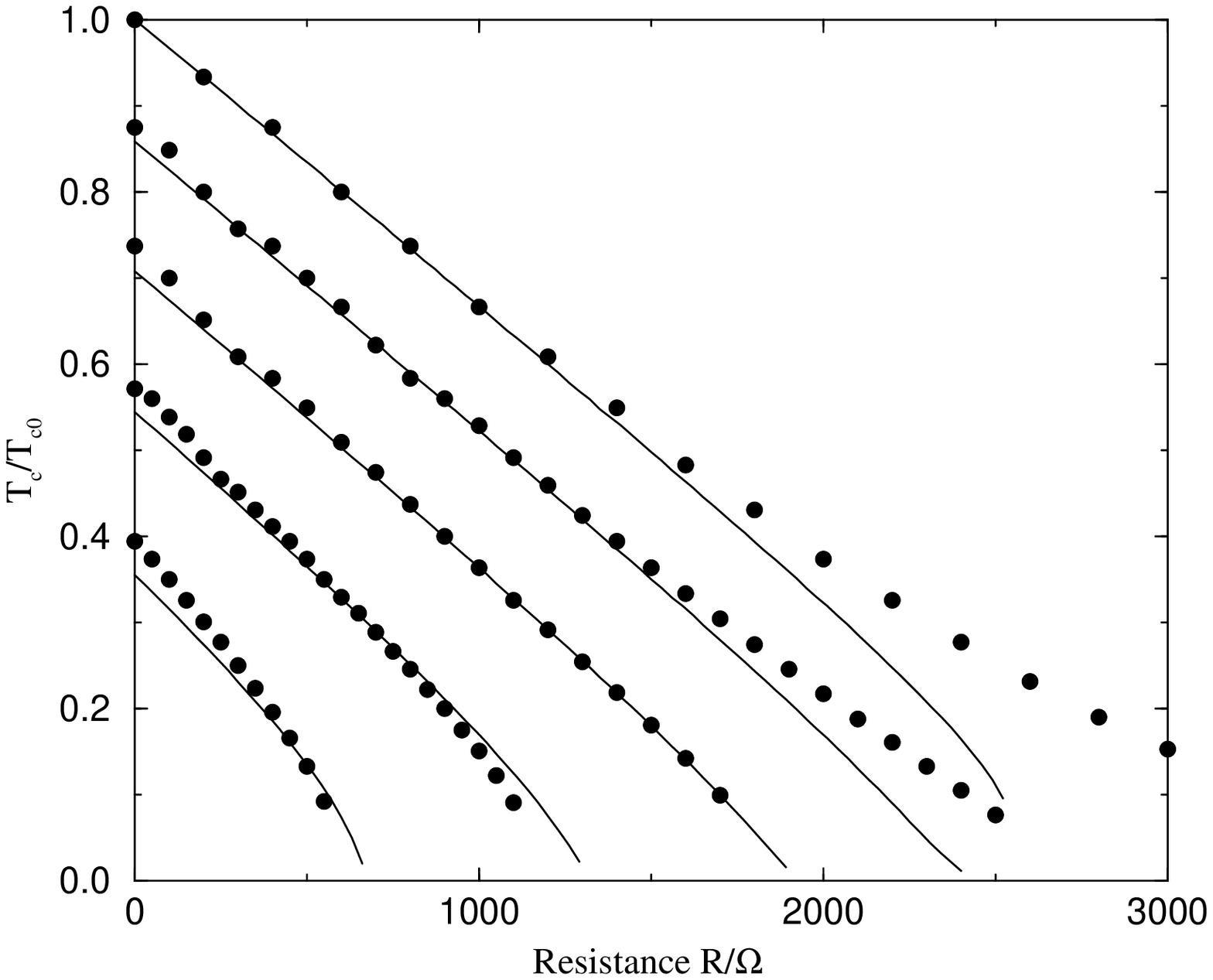,width=9cm}\hskip 0.25truein
\psfig{figure=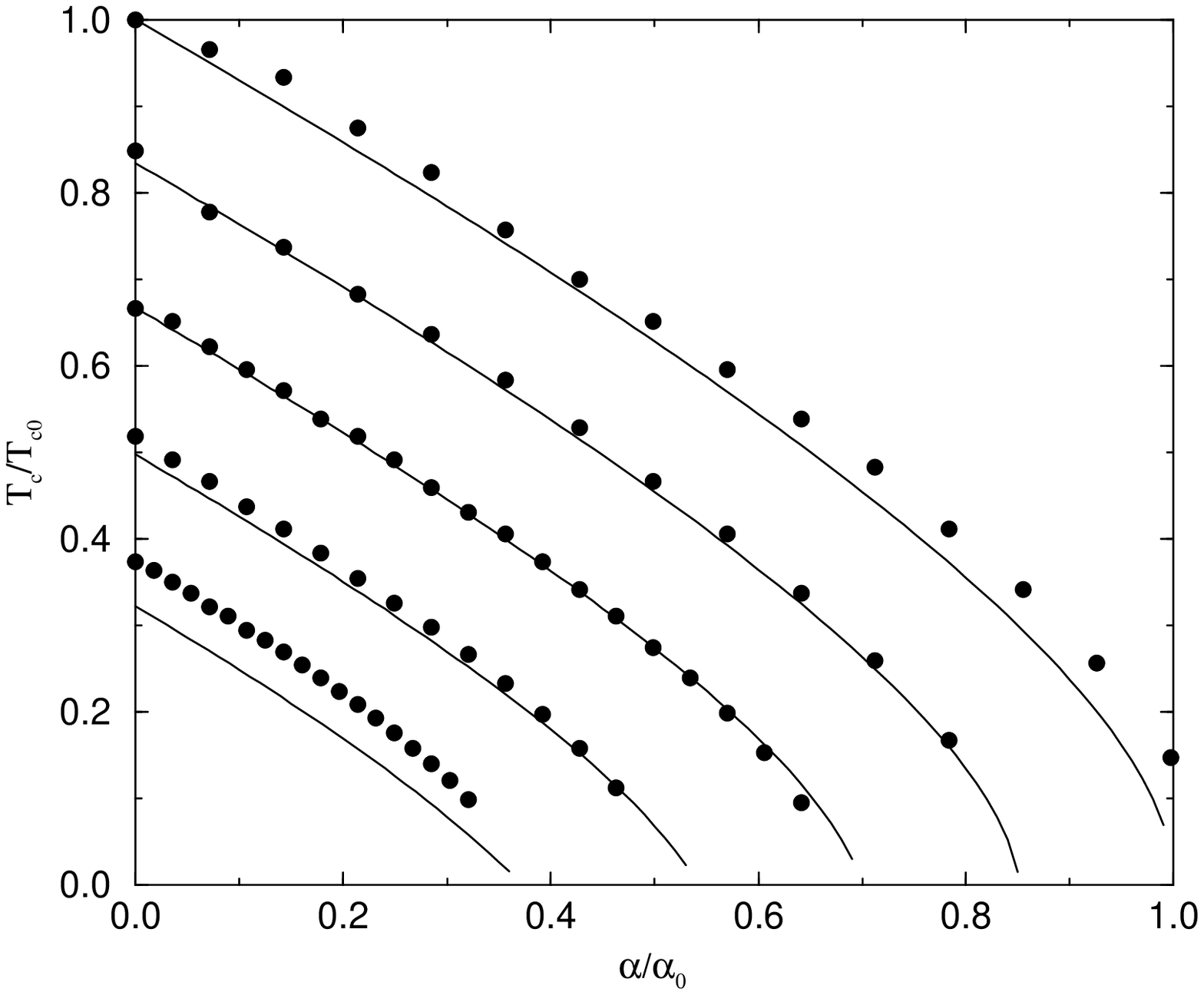,width=9cm}}
\medskip
\caption{Plots of transition temperature as a function of resistance
per square and spin-flip scattering rate. The plot on the left shows
$T_c$ as a function of $R_{\square}$ for values of (top to bottom)
$\alpha=1/2\pi T_{co}\tau_s$ equal to $0$, $0.2$, $0.4$, $0.6$ and $0.8$
times the critical value $\alpha_0$. We see that the $\alpha=0$ curve
has a re-entrance problem, but that the situation improves for finite
$\alpha$. The circles are the results from the non-perturbative resummation
technique. We see that they are roughly in agreement with the perturbation
theory. The plot on the right is of $T_c$ as a function of spin-flip
scattering measured in the dimensionless form $\alpha/\alpha_0$ for values
of $R_{\square}$ equal to (top to bottom)
$0\Om$, $500\Om$, $1000\Om$, $1500\Om$ and $2000\Om$.
The circles are the non-perturbative resummation results, and are again
in good agreement with perturbation theory except for the $2000\Om$ curve.
This might be expected since this curve is very close to the
superconductor-insulator transition.}
\end{figure}

\newpage

\begin{figure}
\centerline{\psfig{figure=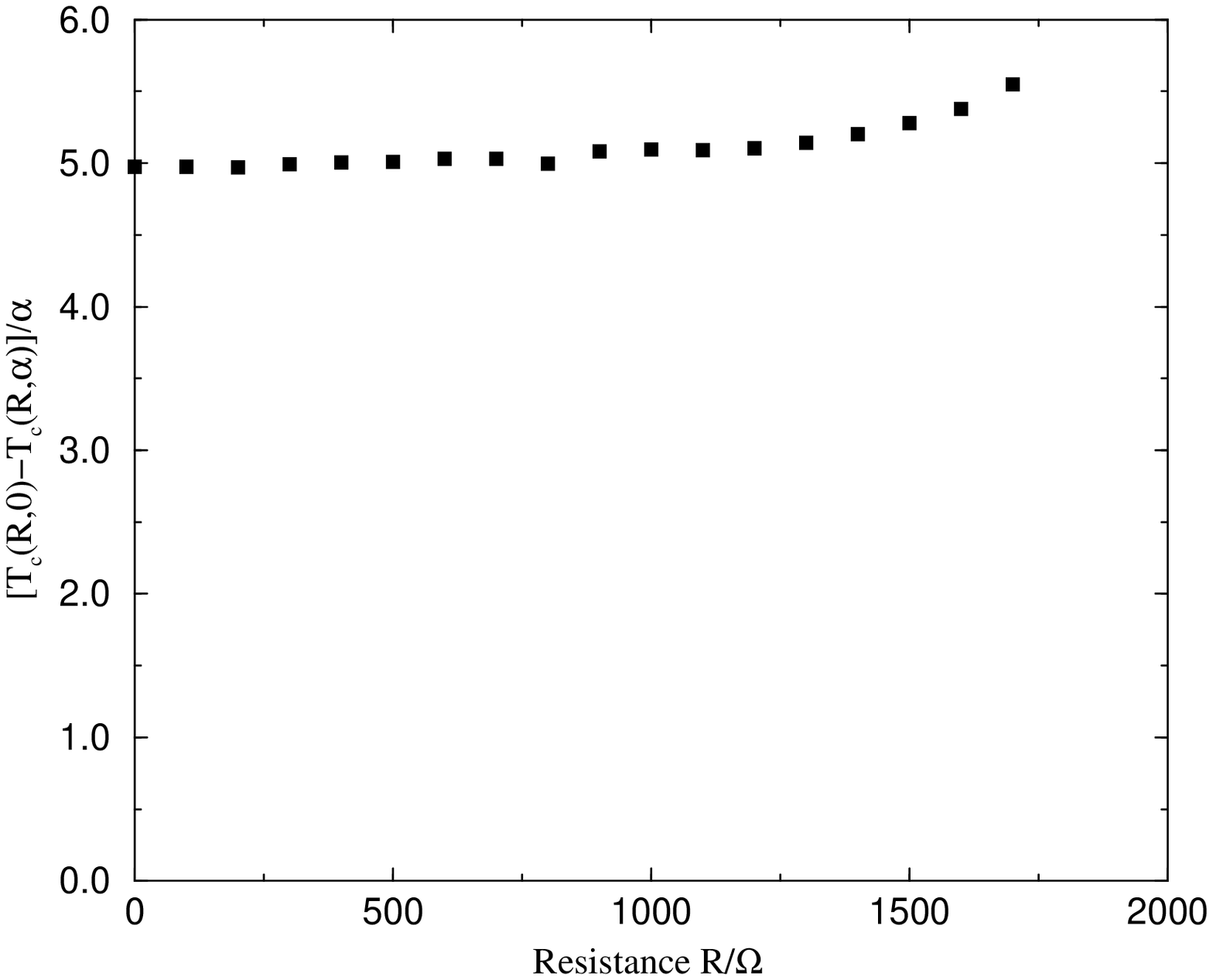,width=8cm}\hskip 0.25truein
\psfig{figure=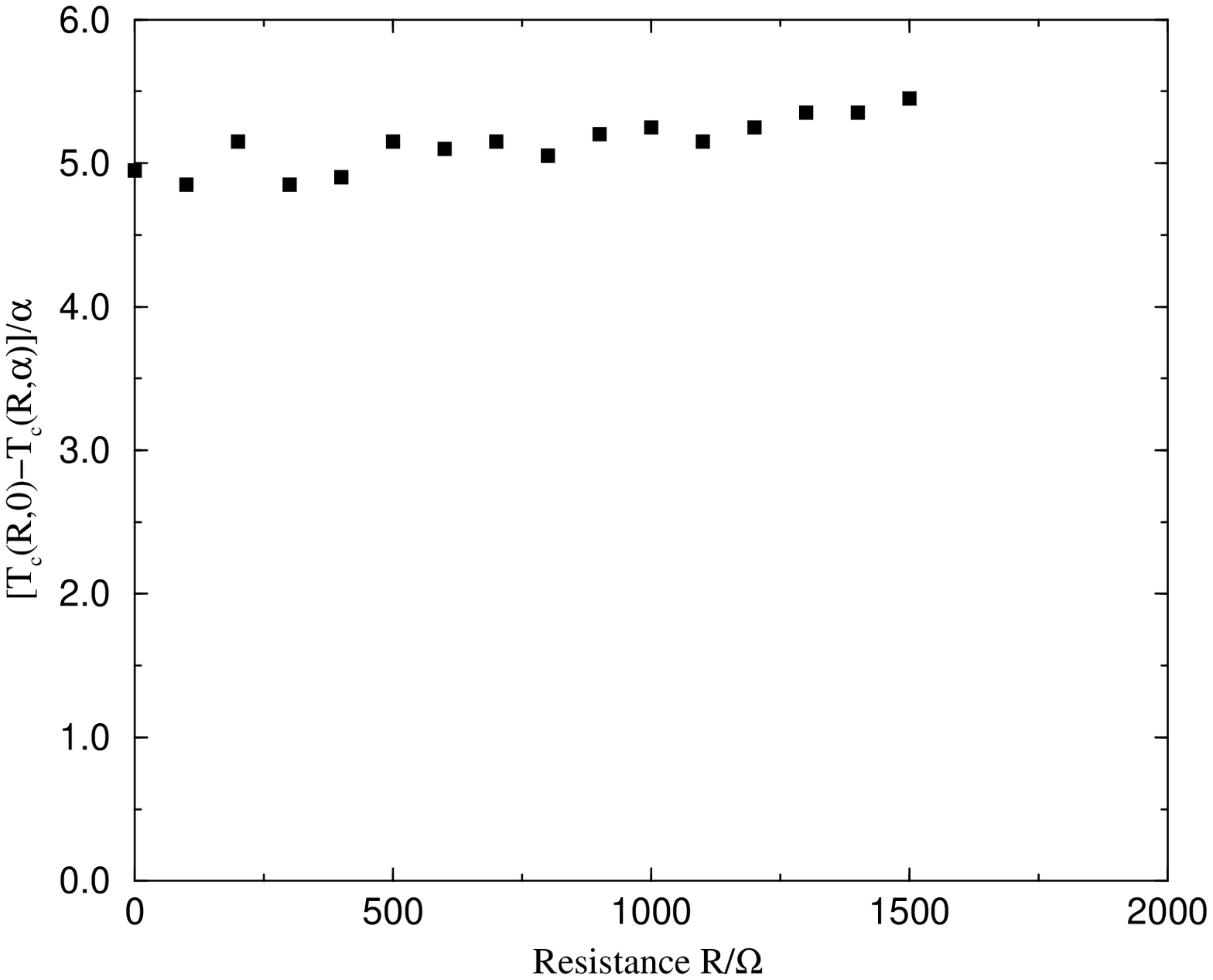,width=8cm}}
\medskip
\caption{Plot of pair-breaking rate per impurity versus resistance
per square of film. We see that this is roughly constant, only increasing
very near to the superconductor-insulator transition, with a variation of
only 10\% over the whole range. The curve on the left is from perturbation
theory; the curve on the right from the non-perturbative resummation.}
\end{figure}


\begin{references}

\bibitem{Fink94} A useful review of the whole area can be found in:
A.M. Finkel'stein, Physica {\bf 197B}, 636 (1994).

\bibitem{RLCM} H.R. Raffy, R.B. Laibowitz, P. Chaudhari and S. Maekawa,
Phys. Rev. B {\bf 26}, 6607 (1983).

\bibitem{GB} J.M. Graybeal and M.R. Beasley,  Phys. Rev. B {\bf 29},
4167 (1984).

\bibitem{HLG} D.B. Haviland, Y. Liu and A.M. Goldman, Phys. Rev. Lett.
{\bf 62}, 2180 (1989).

\bibitem{LK} S.J. Lee and J.B. Ketterson, Phys. Rev. Lett. {\bf 64},
3078 (1990)

\bibitem{HP} A.F. Hebard and M.A. Paalanen, Phys. Rev. B {\bf 30},
4063 (1984).

\bibitem{OKOK} S. Okuma, F. Komori, Y. Ootuka and S. Kobayashi,
J. Phys. Soc. Jpn. {\bf 52}, 3269 (1983).

\bibitem{VDG} J.M. Valles Jnr, R.C. Dynes and J.P. Garno, Phys. Rev. B
{\bf 40}, 6680 (1989); Phys. Rev. Lett. {\bf 69}, 3567 (1992).
                                                                        
\bibitem{Vall94} J.M. Valles Jnr, S-Y Hsu, R.C. Dynes and J.P. Garno,
Physica {\bf 197B}, 522 (1994).

\bibitem{Fink87} A.M. Finkel'stein, Pis'ma Zh. Eksp. Teor. Fiz. {\bf 45}, 37
(1987) [JETP Lett. {\bf 45}, 46 (1987)].

\bibitem{SRW} R.A. Smith, M.Y. Reizer and J.W. Wilkins, Phys. Rev. B
{\bf 51}, 6470 (1995).

\bibitem{OF} Y. Oreg and A.M. Finkel'stein, Phys. Rev. Lett.
{\bf 83}, 191 (1999).

\bibitem{Belitz} D. Belitz, Phys. Rev. B {\bf 35}, 1636 (1987); {\bf 35},
1651 (1987); {\bf 40}, 111 (1989).

\bibitem{CV} J.A. Chervenak and J.M. Valles Jnr, Phys. Rev. B
{\bf 51}, 11977 (1995).

\bibitem{Grif} V. Ambegaokar and A. Griffin, Phys. Rev.
{\bf 137}, A1151 (1964), Appendix A.

\bibitem{PWA} P.W. Anderson, J. Phys. Chem. Solids {\bf 11}, 26 (1959).

\bibitem{AG} A.A. Abrikosov and L.P. Gor'kov, Zh. Eksp. Teor. Fiz. {\bf 39},
1781 (1961) [Sov. Phys. JETP {\bf 12}, 1243 (1961)].

\bibitem{EP88} U. Eckern and F. Pelzer, J. Low. Temp. Phys. {\bf 73},
433 (1988).

\bibitem{MWH} N.D. Mermin and H. Wagner, Phys. Rev. Lett. {\bf 17},
1133 (1966); P.C. Hohenberg, Phys. Rev. {\bf 158}, 383 (1967).

\bibitem{Fink83} A.M. Finkel'stein, Zh. Eksp. Teor. Fiz. {\bf 84}, 168 (1983)
[Sov. Phys. JETP {\bf 57}, 97 (1983)]; Z. Phys. B: Condens. Matter {\bf 56},
189 (1984).

\bibitem{Dev96} T.P. Devereaux and D. Belitz, Phys. Rev. B {\bf 53}, 359 (1996)
.

\bibitem{Namb60} Y. Nambu, Phys. Rev. {\bf 117}, 648 (1960).

\bibitem{Amb61} V. Ambegaokar and L.P. Kadanoff, Nuovo Cimento {\bf 22},
914 (1961).

\bibitem{Schr} J.R. Schrieffer, {\it Theory of Superconductivity}
(Perseus, Reading MA, 1964), Chap 8.

\bibitem{Kad61} G. Baym and L.P. Kadanoff, Phys. Rev. {\bf 124}, 287 (1961).

\bibitem{PWA58} P.W. Anderson, Phys. Rev. {\bf 112}, 1900 (1958).

\bibitem{Rick58} G. Rickayzen, Phys. Rev. {\bf 111}, 817 (1958).


\end{references}
\end{document}